\documentclass[twocolumn,showpacs,aps,prb]{revtex4}
\usepackage{amsmath}
\usepackage{amssymb}
\usepackage{graphicx}
\usepackage{color}

\newcommand{\be}{\begin{equation}}
\newcommand{\ee}{\end{equation}}
\newcommand{\bea}{\begin{eqnarray}}
\newcommand{\eea}{\end{eqnarray}}
\newcommand{\bse}{\begin{subequations}}
\newcommand{\ese}{\end{subequations}}
\newcommand{\sma}{${\rm SrMn_2As_2}$}
\newcommand{\bma}{${\rm BaMn_2As_2}$}
\newcommand{\cma}{${\rm CaMn_2As_2}$}

\newcommand{\cas}{${\rm CaAl_2Si_2}$}
\newcommand{\tcs}{${\rm ThCr_2Si_2}$}
\newcommand{\scma}{${\rm SrMn_2As_2}$ and ${\rm CaMn_2As_2}$}

\begin{document}

\title{Strong magnetic correlations to 900~K in single crystals of the trigonal antiferromagnetic insulators SrMn$_2$As$_2$ and CaMn$_2$As$_2$}

\author{N. S. Sangeetha}
\author{Abhishek Pandey}
\altaffiliation{Present address: Department of Physics and Astronomy, Texas A\&M University, College Station, Texas 77840-4242, USA}
\author{Zackery A. Benson}
\author{D. C. Johnston}
\email{johnston@ameslab.gov}

\affiliation{Ames Laboratory and Department of Physics and Astronomy, Iowa State University, Ames, Iowa 50011, USA}

\date{\today}

\begin{abstract}

Crystallographic, electronic transport, thermal and magnetic properties are reported for \sma\ and \cma\ single crystals grown using Sn flux. Rietveld refinements of powder x-ray diffraction data show that the two compounds are isostructural and crystallize in the trigonal \cas-type structure (space group $P\bar{3}m1$), in agreement with the literature.  Electrical resistivity $\rho$ versus temperature~$T$ measurements demonstrate insulating ground states for both compounds with activation energies of 85~meV for \sma\ and 61~meV for \cma.  In a local-moment picture, the Mn$^{+2}$ $3d^5$ ions are expected to have high-spin $S=5/2$ with spectroscopic splitting factor $g\approx2$.  Magnetic susceptibility $\chi$ and heat capacity measurements versus~$T$ reveal antiferromagnetic (AFM) transitions at $T{\rm_N}= 120(2)$~K and 62(3)~K for \sma\ and \cma, respectively.  The anisotropic $\chi(T\leq T{\rm_N})$ data indicate that the hexagonal $c$~axis is the hard axis and hence that the ordered Mn moments are aligned in the $ab$~plane. The $\chi(T)$ data for both compounds and the $C_{\rm p}(T)$ for \sma\ show strong dynamic short-range AFM correlations from $T_{\rm N}$ up to at least 900~K, likely associated with quasi-two-dimensional connectivity of strong AFM exchange interactions between the Mn spins within the corrugated honeycomb Mn layers parallel to the $ab$~plane.

\end{abstract}


\maketitle

\section{Introduction}

The body-centered tetragonal $AM_2X_2$ ternary compounds ($A$ = rare or alkaline earth, $M$ = transition metal, $X$ = Si, Ge, P, As, Sb) with the ${\rm ThCr_2Si_2}$ structure\cite{Just1996} have generated tremendous interest in the scientific community due to their novel electronic and magnetic properties. Prominent among these is the iron-arsenide family of parent compounds $A{\rm Fe_2As_2}$ ($A$ = Ca, Sr, Ba, Eu).\cite{Johnston2010, Stewart2011, Scalapino2012, Dagotto2013, Fernandes2014, Hosono2015, Dai2015, Inosov2016, Si2016} These materials are metallic and show nearly contiguous  antiferromagnetic (AFM) spin-density wave and structural transitions at temperatures~$T$ up to \mbox{$\sim 200$~K}.   The suppression of these transitions by external pressure or chemical doping leads to superconductivity with bulk superconducting transition temperatures $T_{\rm c}$ up to 56~K\@. It is believed that the FeAs layer as the conducting sheet in this structure plays a crucial role in the occurrence of superconductivity.\cite{Johnston2010, Stewart2011, Scalapino2012, Dagotto2013, Fernandes2014, Hosono2015, Dai2015, Inosov2016, Si2016}  Hence, it is important to investigate other related materials with similar compositions and structures in the search for new superconductors and other novel phenomena. For example, SrNi$_2$As$_2$ ($T_{\rm c} = 0.62$~K, Ref.~\onlinecite{Bauer2008}) and BaNi$_2$As$_2$ ($T_{\rm c} = 0.7$~K, Ref.~\onlinecite{Ronning2008}) were both found to be superconductors.

On the other hand, SrCo$_2$As$_2$ (Refs.~\onlinecite{Pandey2013, Jayasekara2013, Wiecki2015}) and BaCo$_2$As$_2$ (Refs.~\onlinecite{Sefat2009, Anand2014}) are correlated metals with no structural, superconducting or long-range magnetic ordering transitions.  From inelastic neutron scattering measurements, SrCo$_2$As$_2$ is found to exhibit strong AFM correlations at the same stripe wavevector as do the superconducting iron arsenides, which raises the interesting question of why SrCo$_2$As$_2$ is not a high-$T_{\rm c}$ superconductor.\cite{Jayasekara2013}  The reason has been suggested from NMR measurements to be that SrCo$_2$As$_2$ exhibits strong {\it ferromagnetic} (FM) spin correlations/fluctuations in addition to the AFM correlations and these compete with the AFM correlations that are the presumptive glue for superconductivity in these systems.  Subsequent NMR studies indicated that the large range of $T_{\rm c}$ observed within the FeAs-based systems may also arise from the competition between FM and AFM correlations.\cite{Wiecki2015b}

Recently significant attention has focussed on Mn arsenides. Our studies of the properties of the parent and doped BaMn$_2$As$_2$ systems were originally motivated by their potential to be ThCr$_2$Si$_2$-type high-$T_{\rm c}$ superconductors analogous to the cuprates.  The semiconductor BaMn$_2$As$_2$ shows G-type (checkerboard-type) local-moment collinear AFM order below its high N\'eel temperature $T_{\rm N} = 625$~K with the ordered moments aligned along the tetragonal $c$~axis.\cite{Singh2009, Singh2009b, Johnston2011} Thus magnetoelastic coupling does not cause a distortion of the crystal structure below $T_{\rm N}$, contrary to the orthorhombic distortion associated with AFM ordering in the $A{\rm Fe_2As_2}$ compounds due to the collinear Fe ordered moments aligned in the $ab$~plane.   An optical gap of 48~meV was inferred for BaMn$_2$As$_2$ from the optical conductivity,\cite{Antel2012} consistent with results from the electrical resistivity $\rho$ versus temperature~$T$ measurements in the $ab$~plane.\cite{Singh2009}  Furthermore, this optical study\cite{Antel2012} found that BaMn$_2$As$_2$ is much more two-dimensional in its electronic properties than are the $A{\rm Fe_2As_2}$ parent compounds.\cite{Singh2009}  A neutron scattering study of isostructural ${\rm BaMn_2Bi_2}$ found the same G-type AFM structure as in ${\rm BaMn_2As_2}$ but with a lower $T_{\rm N} = 387$~K.\cite{Calder2014}

Only 1.6\% K substitution for Ba transforms BaMn$_2$As$_2$ into a local-moment AFM metal.\cite{Pandey2012, Yeninas2013} Higher doping levels lead to the onset of FM at $\approx 16$\% K-doping\cite{Bao2012} and half-metal FM behavior below the Curie temperature $T_{\rm C} \sim 100$~K at 40\% K~doping (Refs.~\onlinecite{Pandey2013b, Ueland2015}) and at 60\% Rb doping.\cite{Pandey2015}  The FM is thus thought to be associated with FM ordering of the itinerant doped-hole spins  and coexists with the G-type AFM order of the local Mn moments with $T_{\rm N}>300$~K.\cite{Ueland2015,Lamsal2013} 

Unlike BaMn$_2$As$_2$ with the tetragonal ${\rm ThCr_2Si_2}$ structure, the compounds \sma\ and \cma\ both crystallize in the trigonal \cas-type structure\cite{Mewis1978, Brechtel1978} containing a corrugated honeycomb Mn sublattice which can be viewed as a triangular lattice bilayer. The possibility of geometrically-frustrated triangular-lattice exchange connectivity exists and such compounds often show novel physical behaviors associated with the geometric frustration.\cite{Ramirez1994, Moessner2006, Balents2010}  Single crystals of \sma\ were grown previously using Sn flux.\cite{Wang2011}  These authors' in-plane electrical resistivity $\rho(T)$ measurements indicated that the ground state is insulating with activation energies of 0.29--0.64~eV depending on the $T$ range, and their magnetic susceptibility $\chi(T)$ measurements indicated an AFM transition at $T_{\rm N} = 125$~K\@.\cite{Wang2011}

Two neutron powder diffraction studies\cite{Ratcliff2009, Bridges2009} of the related \cas-type ${\rm CaMn_2Sb_2}$ revealed AFM ordering below $T_{\rm N}= 88$~K and 85~K, respectively, with an AFM propagation vector ${\bf k} = (0,0,0)$, i.e., the crystal and AFM unit cells are the same.  In the former paper the AFM structure was deduced to be collinear, with the ordered moments aligned in the $ab$~plane with a low-$T$ ordered moment of 2.8(1)~$\mu_{\rm B}$/Mn, where $\mu_{\rm B}$ is the Bohr magneton.  In the latter paper, a model was favored with the ordered moments canted at $\pm 25^\circ$ with respect to the $ab$ plane with an ordered moment of 3.38(6)~$\mu_{\rm B}$/Mn.

Herein, we report the growth, crystal structure, $\rho(T)$, magnetization as a function of magnetic field $M(H)$, $\chi(T)$ and heat capacity $C_{{\rm p}}(T)$ measurements of \cma\ and \sma\ single crystals.  These studies were initiated because of the above-noted possibility that the Mn spin lattice might exhibit novel magnetic behaviors associated with the presence of geometric frustration within the triangular-lattice Mn layers.  If the strongest AFM interactions are indeed within a triangular lattice layer, this should lead to a noncollinear AFM structure below $T_{\rm N}$.  Instead, in a companion neutron diffraction study to the present work, the AFM structure of \sma\ was found to be collinear with the ordered Mn moments aligned in the $ab$ plane with magnitude 3.6~$\mu_{\rm B}$/Mn.\cite{Das2016}  This magnetic structure is the same as one of the two AFM structures proposed for ${\rm CaMn_2Sb_2}$ (Ref.~\onlinecite{Ratcliff2009}) discussed above.

We discovered that the $\chi(T)$ of \sma\ and \cma\ and the $C_{\rm p}(T)$ of \sma\ above their respective N\'eel temperatures $T_{\rm N}$ of 120 and 62~K exhibit behaviors characteristic of strong dynamic short-range AFM spin correlations up to at least 900~K, likely arising from quasi-two-dimensional connectivity of strong AFM Mn--Mn exchange interactions within the corrugated honeycomb Mn spin sublattice. This result is interesting because such strong AFM spin correlations up to high temperatures and the suppression of $T_{\rm N}$ to much lower temperatures than expected from molecular field theory, due to AFM fluctuations associated with the low dimensionality of the exchange interaction connectivity, may give rise to novel physical properties upon doping the compounds into the metallic state.

\section{Experimental Details}

Single crystals of \sma{} and \cma{} were grown using Sn flux. High-purity elements Sr (99.95\%) from Sigma Aldrich, and Ca (99.95\%), Co (99.998\%), As (99.9999\%) and Sn (99.999\%) from Alfa Aesar were taken in the ratio (Sr,Ca):Mn:As:Sn = 1:2:2:20 and placed in an alumina crucible that was subsequently placed in a silica tube that was evacuated, partially refilled with high-purity argon ($\approx$1/4 atm pressure) and then sealed. After preheating at 600$^{\circ}$C for 5 h, the assembly was heated to 1150~$^{\circ}$C at the rate of 50~$^{\circ}$C/h and held at this temperature for 20 h for homogenization. Then the furnace was slowly cooled at the rate of 5~$^{\circ}$C/h to 700~$^{\circ}$C\@. At this temperature the molten Sn flux was decanted using a centrifuge. Shiny hexagonal-shape single crystals of maximum dimensions $4\times 3\times 1\ {\rm mm}^3$ were obtained.

Semiquantitative chemical analyses of the single crystals were performed using a JEOL scanning electron microscope (SEM) equipped with an EDX (energy-dispersive x-ray analysis) detector, where a counting time of 120 s was used. A room-temperature powder x-ray diffraction (XRD) pattern was recorded on crushed single crystals using a Rigaku Geigerflex powder diffractometer with Cu K$\alpha$ radiation at diffraction angles 2$\theta$ from 10$^{\circ}$ to 110$^{\circ}$ with a 0.02$^{\circ}$ step width.  The data were analysed by Rietveld refinement using FullProf software.\cite{fullprof} 

$M(T)$ measurements for ${\rm 1.8~K \leq}~T~{\rm \leq 300~K}$ and $M(H)$ measurements for $H \leq 5.5$~T were carried out using a Quantum Design, Inc., Magnetic Properties Measurement System (MPMS). The high-temperature $M(T)$ for ${\rm 300~K \leq T \leq 900~K}$ was measured using the vibrating sample magnetometer (VSM) option of a Quantum Design, Inc., Physical Properties Measurement System (PPMS)\@.  In this paper we exclusively use Gaussian cgs units for $M$, $\chi$ and $H$ (see Sec.~3.5.1 of Ref.~\onlinecite{Johnston2010}).  In this system of units, the Tesla~(T) is a unit of convenience for~$H$ defined as 1~T~=~$10^4$~Oe, where Oe is the conventional cgs unit for~$H$.

$C{\rm_p}(T)$ data were obtained using a relaxation method with the heat capacity option of the PPMS\@. Four-probe $\rho(T)$ data were obtained with an ac current amplitude $I=1\,\rm{\mu A}$ at a frequency of 37.7~Hz using the ac transport option of the PPMS\@. Electrical contacts to a crystal were made by soldering 0.05~mm diameter Pt wire to a crystal using indium solder.

\section{Experimental Results}

\subsection{\label{Sec:Struct} Crystal Structure}

\begin{figure}
\includegraphics[width=3.5in]{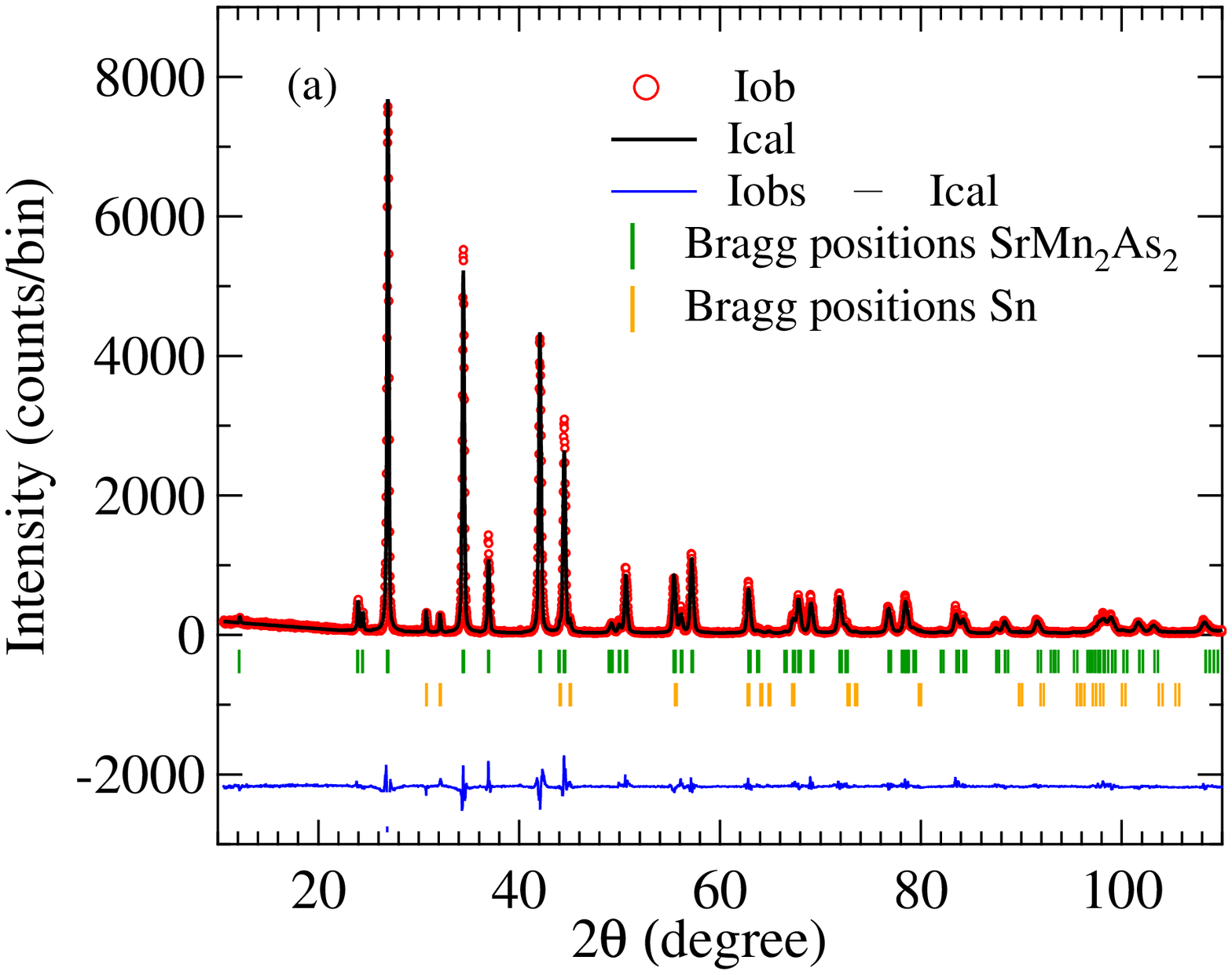} 
\includegraphics[width=3.5in] {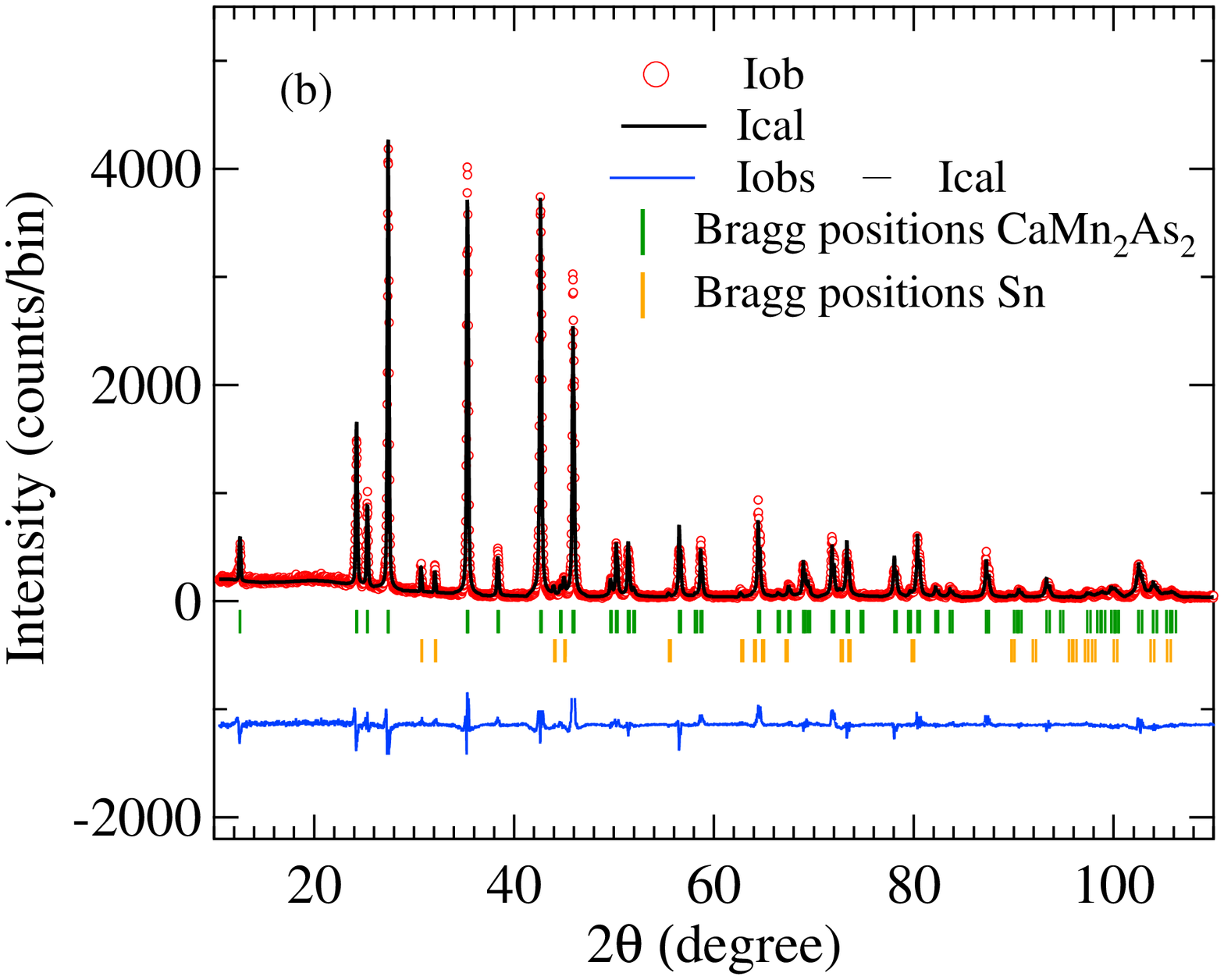}
\caption{\label{Fig:xrd}(Colour online)  Powder x-ray diffraction patterns (open circles) of (a) \sma{} and (b) \cma{} at room temperature. The solid line represents the Rietveld refinement fit calculated for the \cas{}-type trigonal structure with space group $P\bar{3}m1$ together with the Sn impurity phase.}
\end{figure}

\begin{table}
\caption{\label{Table:crystal structure parameter} Crystallographic and Rietveld refinement parameters
obtained from powder XRD of \sma{} and \cma{} crystals. The structures are trigonal ${\rm CaAl_2Si_2}$-type with space group $P\bar{3}m1$. The atomic coordinates of \scma{} are Sr/Ca: $1a$ (0, 0, 0); Mn: $2d$ (1/3, 2/3, $z{\rm_{Mn}}$); and As: $2d$ (1/3, 2/3, $z{\rm_{As}}$). The shortest Mn--Mn interatomic distances in \scma{} [see Fig.~\ref{Fig:crystal_structure}(b)] are also listed.}
\begin{ruledtabular}
\begin{tabular}{lcc}
 				& \sma{} & \cma{}\\
\hline
Lattice parameters  \\
$a$~(\AA)					& 4.2962(1)  	& 4.2376(1)	\\
$c$~(\AA)					& 7.2997(2)  	& 7.0331(2)	\\
$c/a$  					& 1.6991(1)  	& 1.6596(1) 	\\
$V_{\rm cell}~(\rm{\AA}^3)$	& 116.682(6)  	& 109.372(6)	\\
\hline
Atomic coordinates		\\
$z{\rm_{Mn}}$  			& 0.6231(1)  	&  0.6248(4)	\\
$z{\rm_{As}}$   			& 0.2667(2) 	&  0.2537(3) 	\\
\hline
Refinement quality				\\
$\chi^2$					& 3.05  		&  4.03		\\
$R_{\rm p}$ (\%)			& 10.3 		&  12.7 		\\
$R_{\rm wp}$ (\%)			& 13.6 		&  16.4		\\
\hline
Shortest Mn--Mn\\
distances (\AA)\\
$d_1$  					& 3.06306(8)  	&  3.0112(2)\\
$d_2$   					& 4.29620(5) 	&  4.23760(5)\\
$d_3$   					& 5.27633(7) 	&  5.1985(2)\\
$d_{z1}$   				& 6.0357(2) 	&  5.8171(4)\\
$d_{z2}$   				& 7.2997(2) 	&  7.0331(2)\\
\end{tabular}
\end{ruledtabular}
\end{table}

The crystal symmetry of several \scma{} crystals was checked by x-ray Laue back scattering which showed trigonal symmetry with well-defined diffraction spots which clearly indicated the good quality of the crystals. In this paper we use the hexagonal setting for the trigonal unit cell.  The data also revealed that the \scma\ platelike crystals grow with the plate surface parallel to the hexagonal $ab$~plane. SEM imaging and EDX analyses were performed to check the chemical composition and surface morphology of the crystals. The average elemental ratio of the samples was in agreement with the expected 1:2:2 stoichiometry of the compounds to within the errors.  The amount of Sn incorporated into the crystal structure from the Sn flux is zero to within the experimental error. The present analyses did not show any other elements.

The phase purity of our \scma{} crystals was confirmed by powder XRD\@. Their XRD patterns at 300 K along with the results of Rietveld refinements are shown in Figs.~\ref{Fig:xrd}(a) and~\ref{Fig:xrd}(b), respectively. One sees the presence of adventitious elemental Sn flux, so two-phase Rietveld refinements were carried out.  The refinement results confirm that the crystals have the trigonal \cas-type structure with space group $P\bar{3}m1$. The refinement and crystal parameters obtained are listed in Table~\ref{Table:crystal structure parameter}.  The crystal  parameters are in good agreement with previously reported values.\cite{Mewis1978, Brechtel1978, Wang2011}

%__________________Fig:crystal structure____________________________________
\begin{figure*}
\includegraphics[width=2.25in]{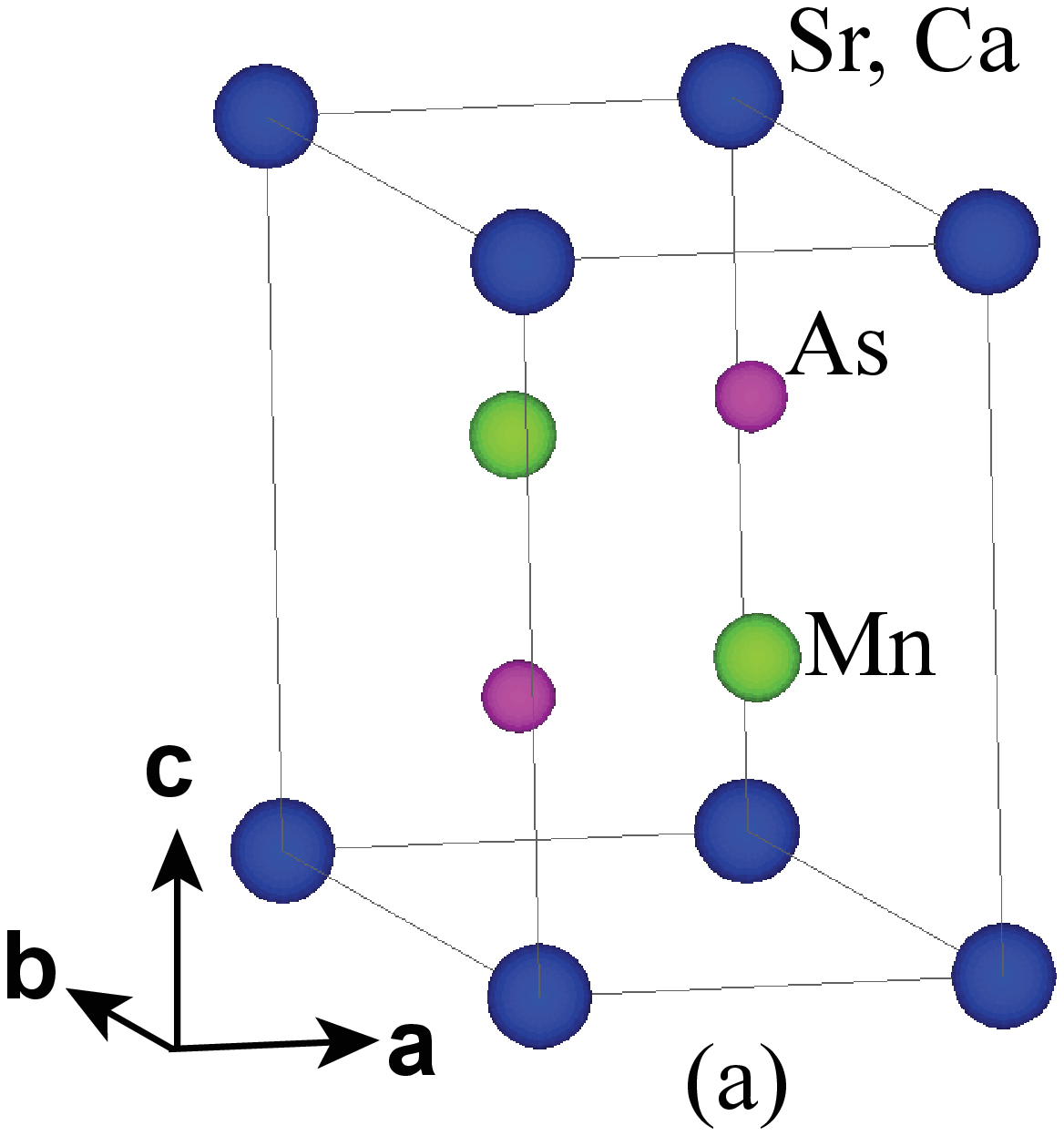}\hspace{0.5in}
\includegraphics[width=2.75in]{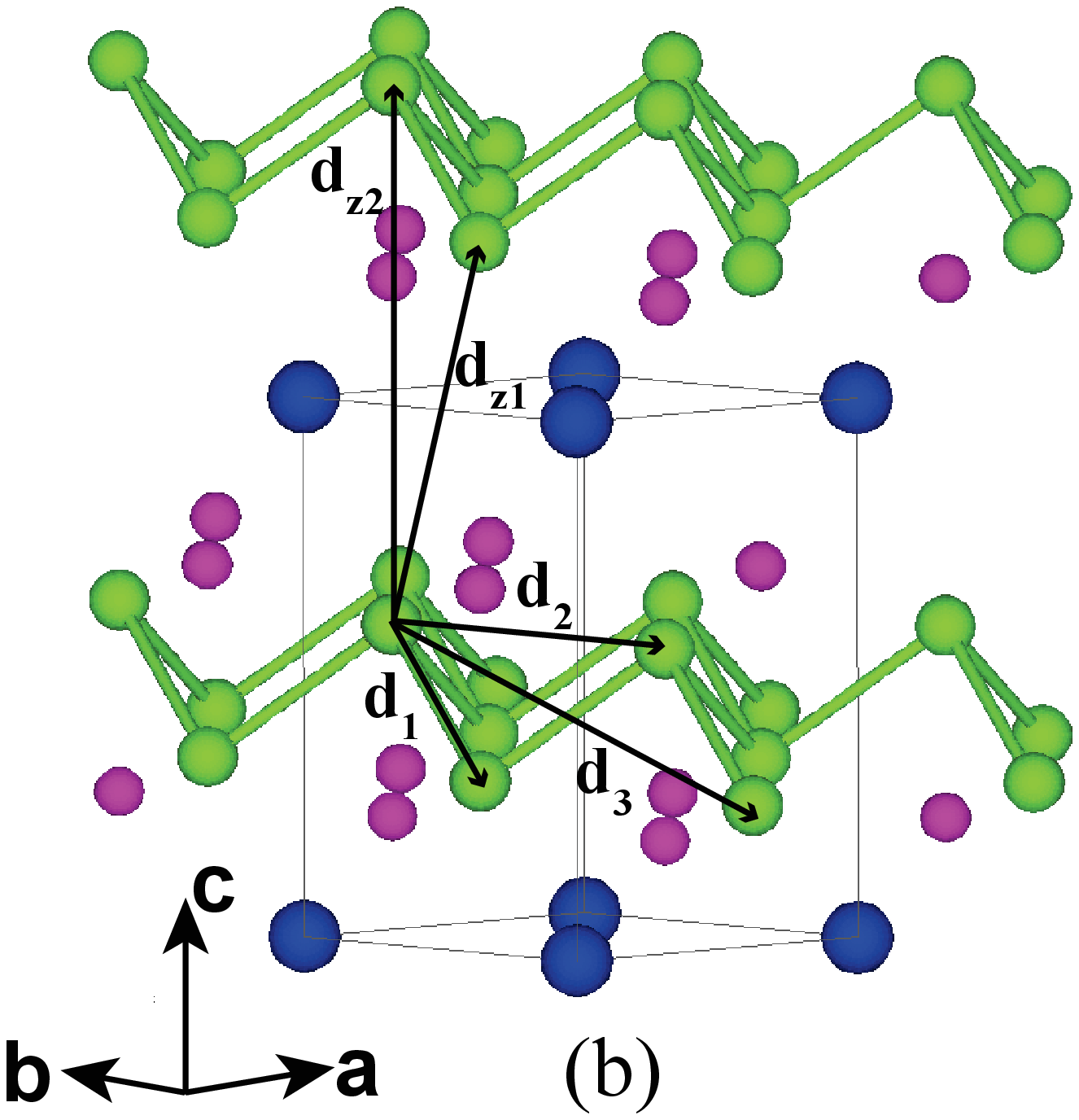}\vspace{0.25in}
\includegraphics[width=2.75in]{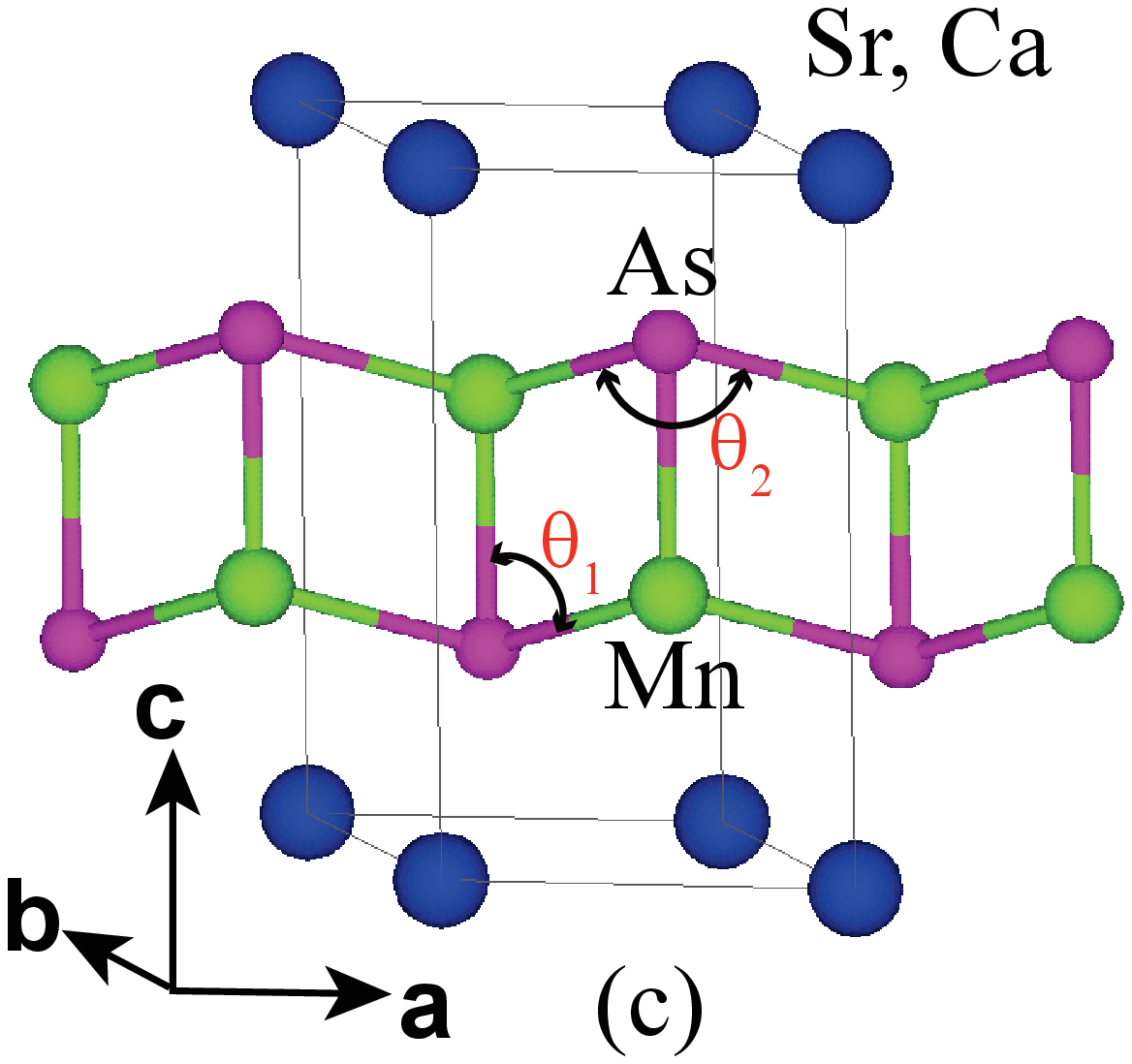}\hspace{0.3in}
\includegraphics[width=2.75in]{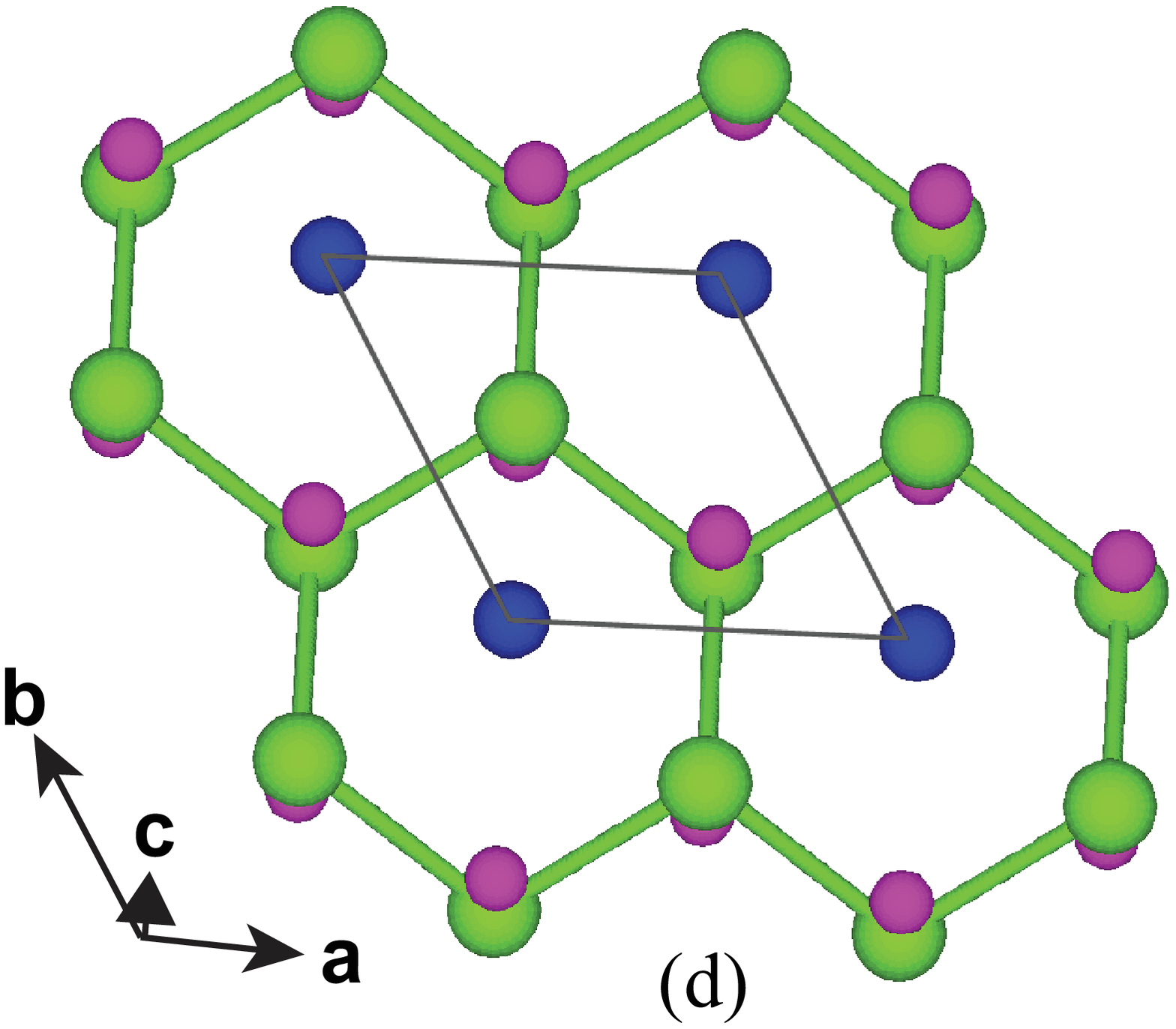}
\caption{(Colour online) Trigonal \cas-type crystal structure of \scma\ in the hexagonal setting.  (a)~Outline of a unit cell containing one formula unit. (b)~Corrugated Mn honeycomb layers as viewed from nearly perpendicular to the $c$~axis. The smallest Mn--Mn interatomic distances within a corrugated Mn honeycomb layer ($d_1$, $d_2$, $d_3$) and between layers ($d_{z1}$, $d_{z2}$) are indicated.  (c)~Expanded view of the structure from a view nearly perpendicular to the $c$~axis showing the corrugated [Mn$_2$As$_2]^{-2}$ honeycomb layers separated by Sr or Ca. (d)~Projection of the Mn sublattice onto the $ab$ plane with a slight $c$ axis component illustrating the corrugated Mn honeycomb lattice. The corrugated honeycomb lattice layer can be viewed as a triangular lattice bilayer [compare with panel~(b)].  The outline of a unit cell in the $ab$~plane is also shown.}
\label{Fig:crystal_structure}
\end{figure*}
%-------------------------------------------------------------------------------------------------------

Figure~\ref{Fig:crystal_structure}(a) shows a unit cell of trigonal \sma{} and \cma{} in the hexagonal setting.  As shown in Figs.~\ref{Fig:crystal_structure}(b) and~\ref{Fig:crystal_structure}(d), the structure consists of  corrugated honeycomb $\left[{\rm Mn_2As_2}\right]^{-2}$ layers that are stacked along the $c$~axis and separated by Sr$^{+2}$ or Ca$^{+2}$ cations, respectively. Alternatively, the Mn sublattice can be viewed as triangular double layers of Mn stacked along the $c$~axis and separated by Ca or Sr atoms.  The three smallest Mn--Mn interatomic distances [see Fig.~\ref{Fig:crystal_structure}(b)] are  within the corrugated Mn honeycomb layers and are listed in Table~\ref{Table:crystal structure parameter}.  The nearest-neighbor Mn--Mn distance ($d_1$) is between the two Mn atoms at different heights ($z$ values) within a unit cell.  The second-nearest-neighbor Mn-Mn distance ($d_2$) is between Mn atoms at the same height in adjacent unit cells along the $ab$~plane forming a triangular-lattice layer, and the third-nearest-neighbor Mn--Mn distance ($d_3$) is between nearest-neighbor Mn atoms in adjacent unit cells in the $ab$~plane.  The nearest- and second-nearest-neighbor distances $d_{z1}$ and $d_{z2}$ between Mn atoms in adjacent layers in different unit cells along the $c$~axis are also listed in Table~\ref{Table:crystal structure parameter}.

Since the minimum intralayer Mn--Mn distance $d_1\approx 3$~\AA\ is much shorter than the minimum interlayer Mn--Mn distance ($d_{z1}\approx 6$~\AA), \cma\ and \sma\ likely have a quasi-two-dimensional Mn--Mn exchange interaction connectivity.  This large spatial anisotropy in the exchange interactions should be obvious from $\chi(T>T_{\rm N})$ measurements, which is confirmed below.  These exchange interactions could arise from direct Mn--Mn interactions and/or from indirect Mn--As--Mn superexchange interactions.  The latter would likely occur via two main paths: (i) between first-nearest-neighbor Mn spins with $\angle{}$Mn-As-Mn ($\theta _1$) = 72$^{\circ}$ and another between second-neighbor Mn spins with $\angle{}$Mn-As-Mn ($\theta _2$) = 111$^{\circ}$ [see Fig.~\ref{Fig:crystal_structure}(c)].  It will be interesting to see which of these interactions are dominant within the unusual trigonal symmetry of the Mn spin lattice.

To summarize, the \scma{} trigonal structure is quite different from the body-centered tetragonal \tcs{} structure found for the $A{\rm Fe_2As_2}$ parent compounds that is composed of metal-arsenide tetrahedra separated by alkaline earth layers. The primary difference between them is the geometry of the transition metal layers.  In \scma, the Mn bilayer is a corrugated Mn honeycomb lattice where each Mn atom is coordinated by three other Mn atoms at $\sim90^{\circ}$ like the corner of the cube as seen in Figs.~\ref{Fig:crystal_structure}(b) and \ref{Fig:crystal_structure}(c), whereas in \tcs-type compounds such as BaMn$_2$As$_2$ or BaFe$_2$As$_2$, the Mn or Fe network is a simple square-planar lattice where each Mn or Fe is coordinated by four other Mn or Fe atoms, also at 90$^{\circ}$ angles between them.  On the other hand, the AFM in \bma\ is quasi-two-dimensional,\cite{Johnston2011} just as we find it to be in \scma\ from the $\chi(T)$ data in Sec.~\ref{Sec:MandChi} below.

\subsection{\label{Sec:Rho} In-Plane Electrical Resistivity}

\begin{figure}
\includegraphics[width=3.5in]{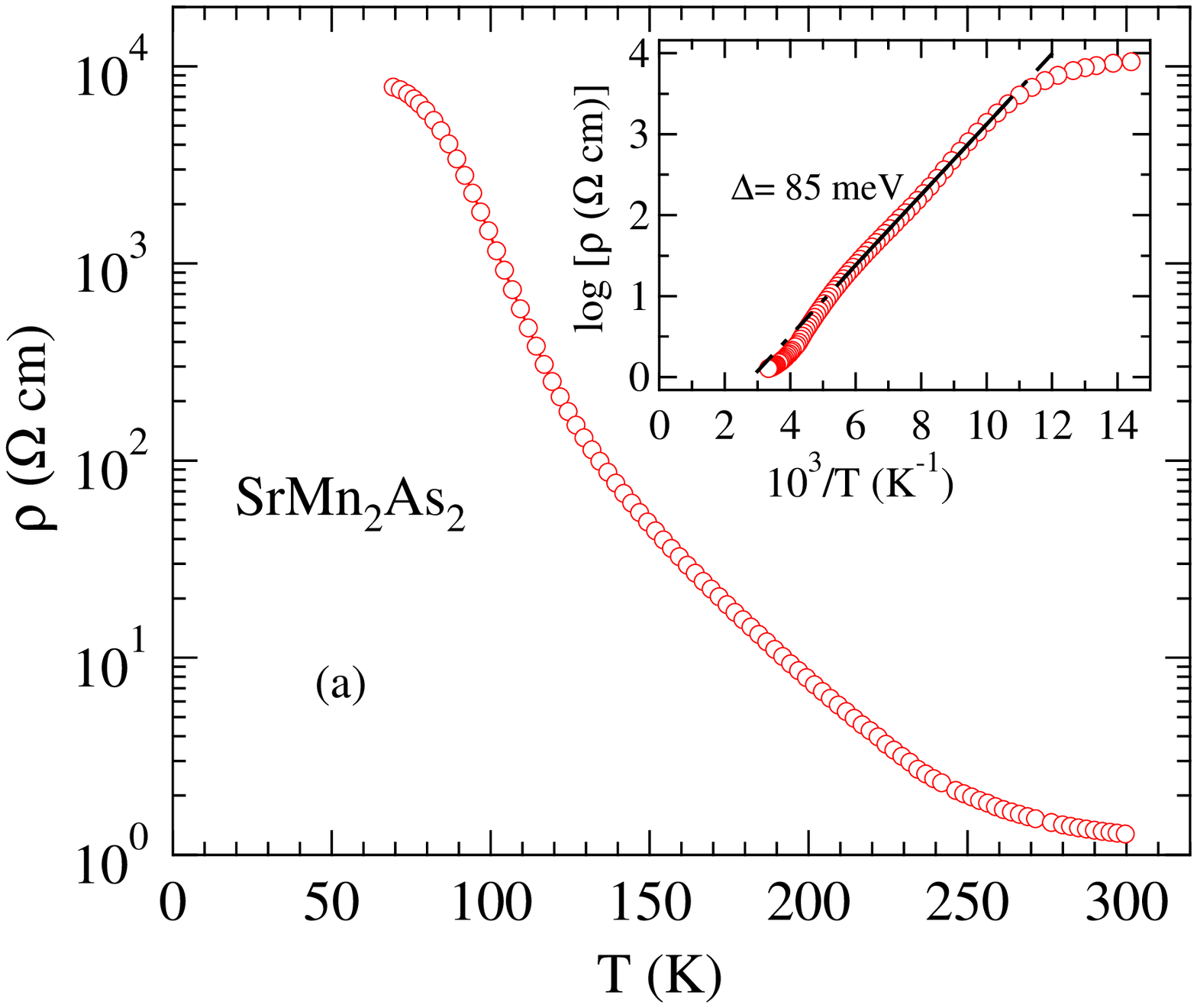}
\includegraphics[width=3.5in]{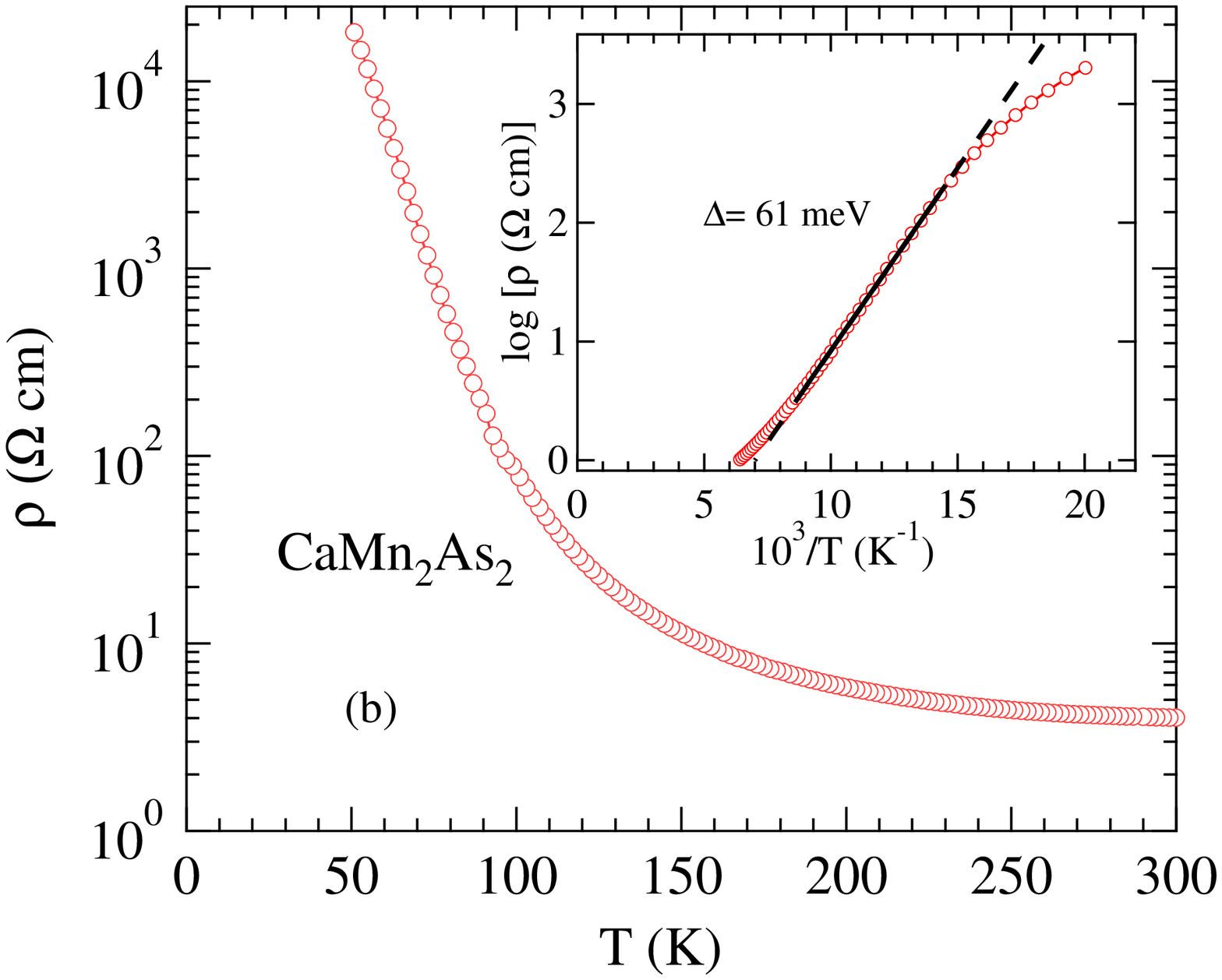}
\caption{(Colour online) In-plane electrical resistivity $\rho$ of (a)~\sma{} and (b) \cma{} versus temperature~$T$ from $\sim 50$ to 300 K\@. The insets show plots of $\log_{10}\rho$ versus 1000/$T$\@. The solid straight lines through the data are fits over the temperature interval between 70 and 120~K by the expression $\log_{10}\rho=A+2.303\Delta/T$ and the resulting fitted activation energies are listed.  The dashed lines are extrapolations.}
\label{Fig:(Sr,Ca)Mn2As2_Res}
\end{figure}

Figures~\ref{Fig:(Sr,Ca)Mn2As2_Res}(a) and \ref{Fig:(Sr,Ca)Mn2As2_Res}(b) show $\rho(T)$ of \sma{} and \cma{}, respectively, from $\sim50$ to 300~K measured in the $ab$ plane.  The data show that both compounds are semiconductors with insulating intrinsic ground states. We fitted $\rho(T)$ in the temperature region between 70 and 120~K by the expression $\log_{10}\rho = A + 2.303\Delta/k_{\rm B}T$, where $A$ is a constant, $k_{\rm B}$ is Boltzmann's constant and $\Delta$ is the activation energy. The fits are shown as the solid straight lines through the data in the insets of Figs.~\ref{Fig:(Sr,Ca)Mn2As2_Res}(a) and~\ref{Fig:(Sr,Ca)Mn2As2_Res}(b), which give the activation energies $\Delta = 85$~meV for \sma{} and $\Delta = 61$~meV for \cma{}. These activation energies are of the same order as previously obtained from $\rho(T)$ data for BaMn$_{2}$As$_{2}$.\cite{Singh2009}  Our activation energy for SrMn$_{2}$As$_{2}$ is significantly smaller than the previously reported values $\Delta = 0.29$ and~0.64~eV, depending on the $T$~range, that were also obtained from single-crystal in-plane $\rho(T)$ data.\cite{Wang2011} 
 
%\newpage

\subsection{\label{Sec:MandChi} Magnetization and Magnetic Susceptibility}

\begin{figure}
\includegraphics[width=3.5in]{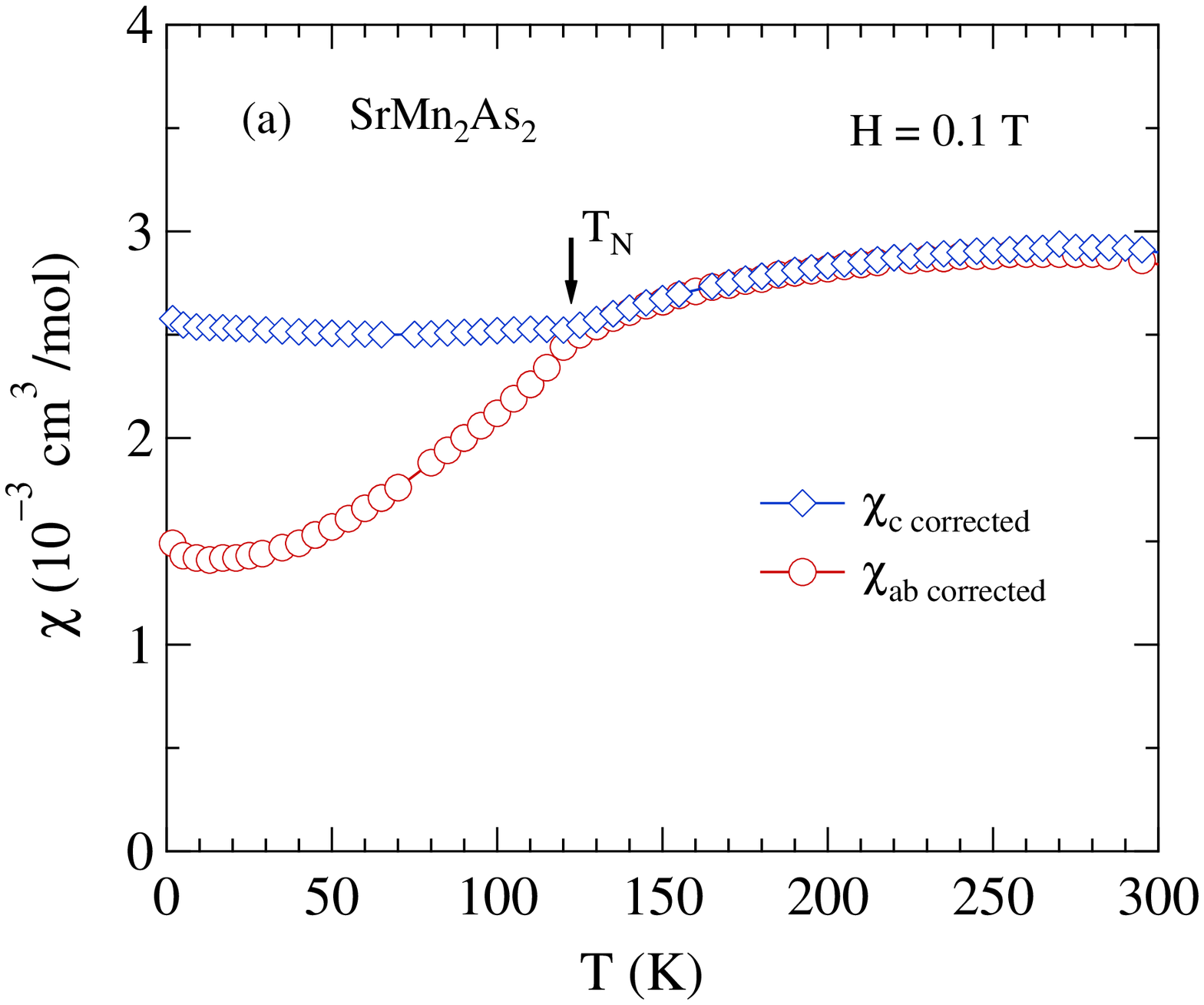}\vspace{-0.2in}
\includegraphics[width=3.5in]{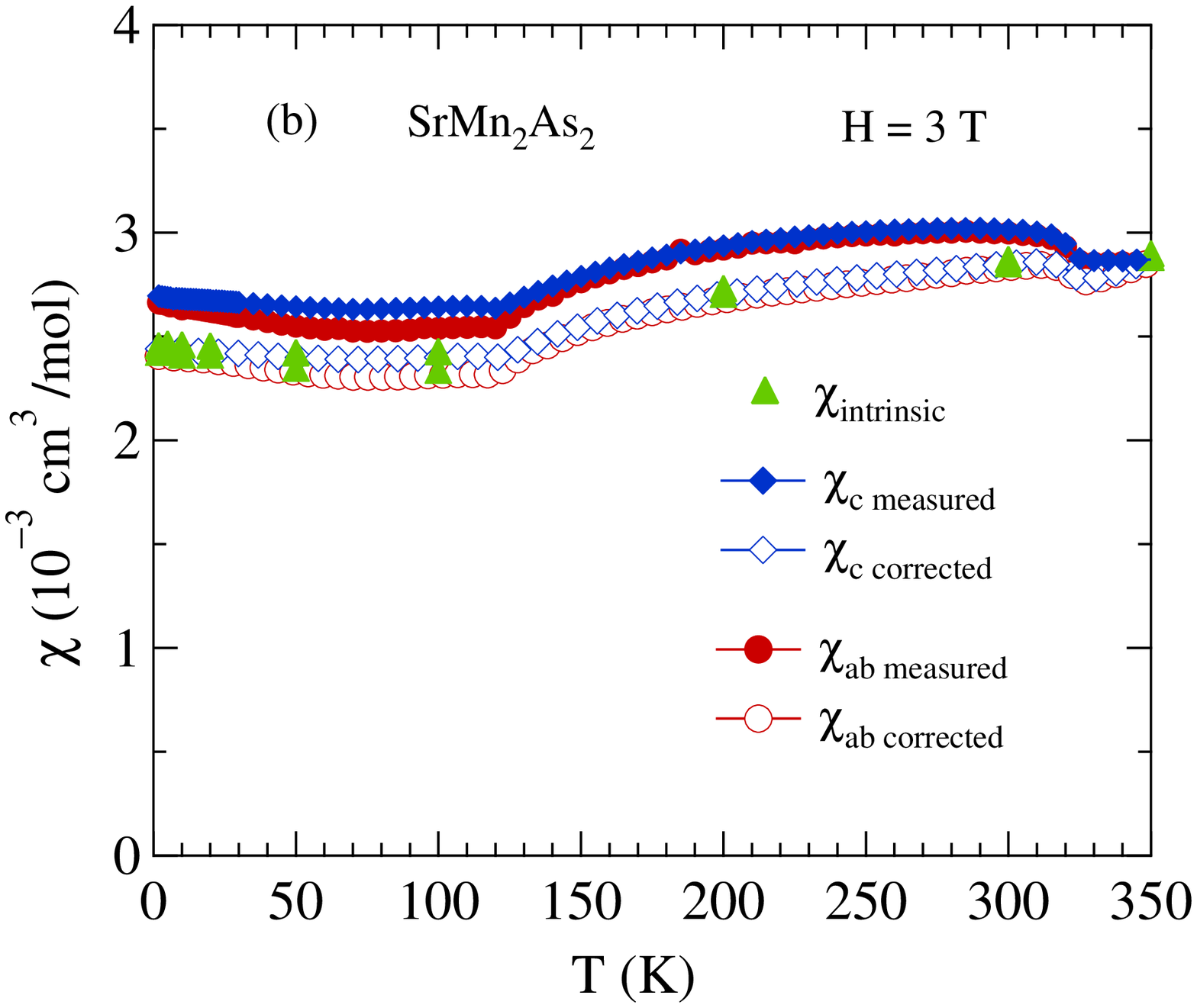}\vspace{-0.2in}
\includegraphics[width=3.5in]{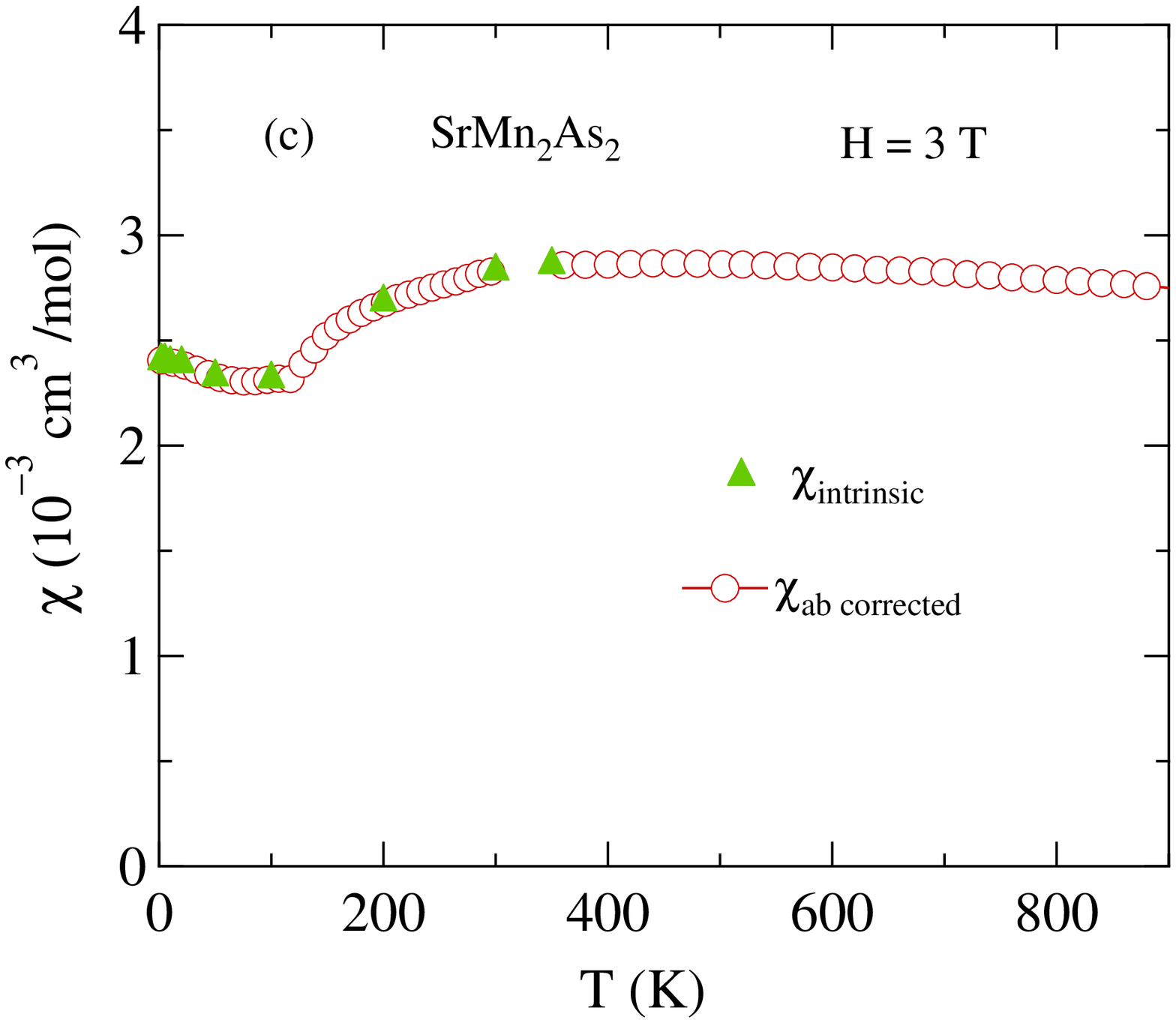}
\caption{(Color online) ZFC magnetic susceptibility $\chi (T)$ of \sma{} from 1.8 to 300 K measured in magnetic fields (a)~$H=0.1$~T and (b)~$H = 3$~T applied in the $ab$~plane ($H\parallel ab,\ \chi_{ab}$) and along the $c$~axis ($H\parallel{c},\ \chi_c$). (c)~ZFC susceptibility versus~$T$ for $1.8\leq T\leq900$~K measured in $H = 3$~T applied in the $ab$~plane ($H\perp c$).  The ``intrinsic'' values are obtained from $M(H)$ isotherms using Eq.~(\ref{eq:1}) and the ``corrected'' ones are obtained from $M(T)$ data using Eq.~(\ref{Eq:Chicorrected}). }
\label{Fig:SrMn2As2_Chi}
\end{figure}

\begin{figure}
\includegraphics[width=3.5in]{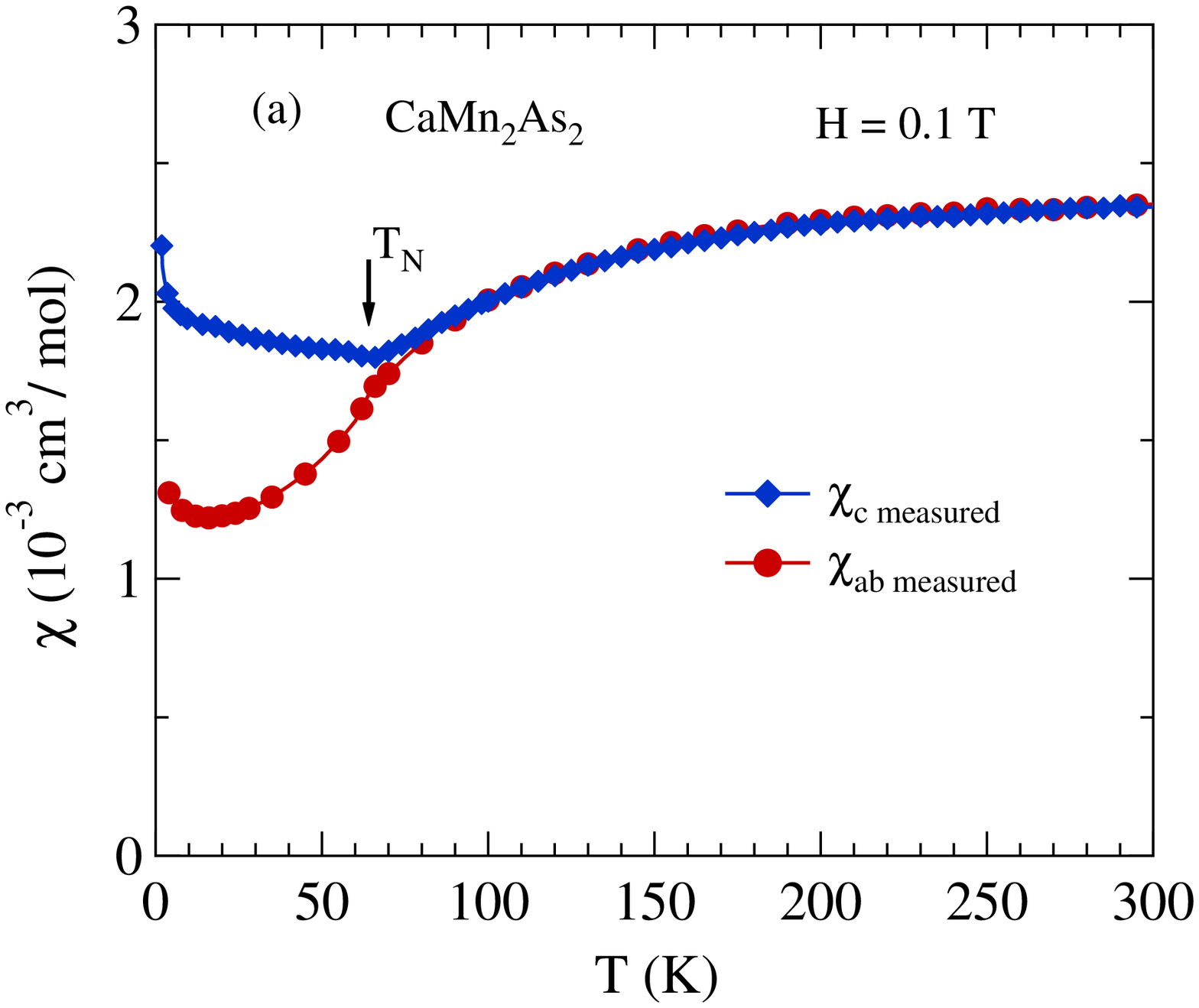}\vspace{-0.2in}
\includegraphics[width=3.5in]{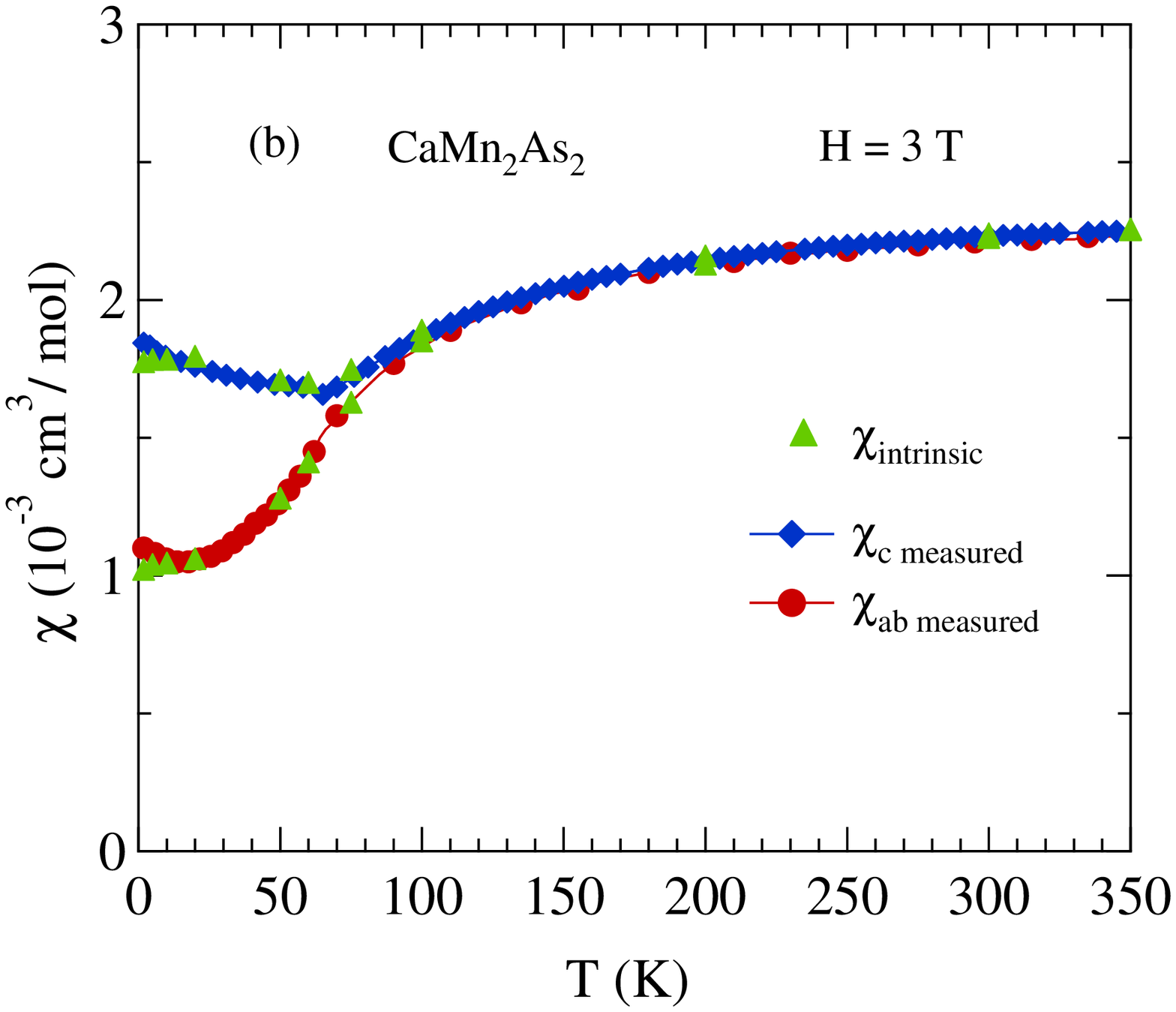}\vspace{-0.2in}
\includegraphics[width=3.5in]{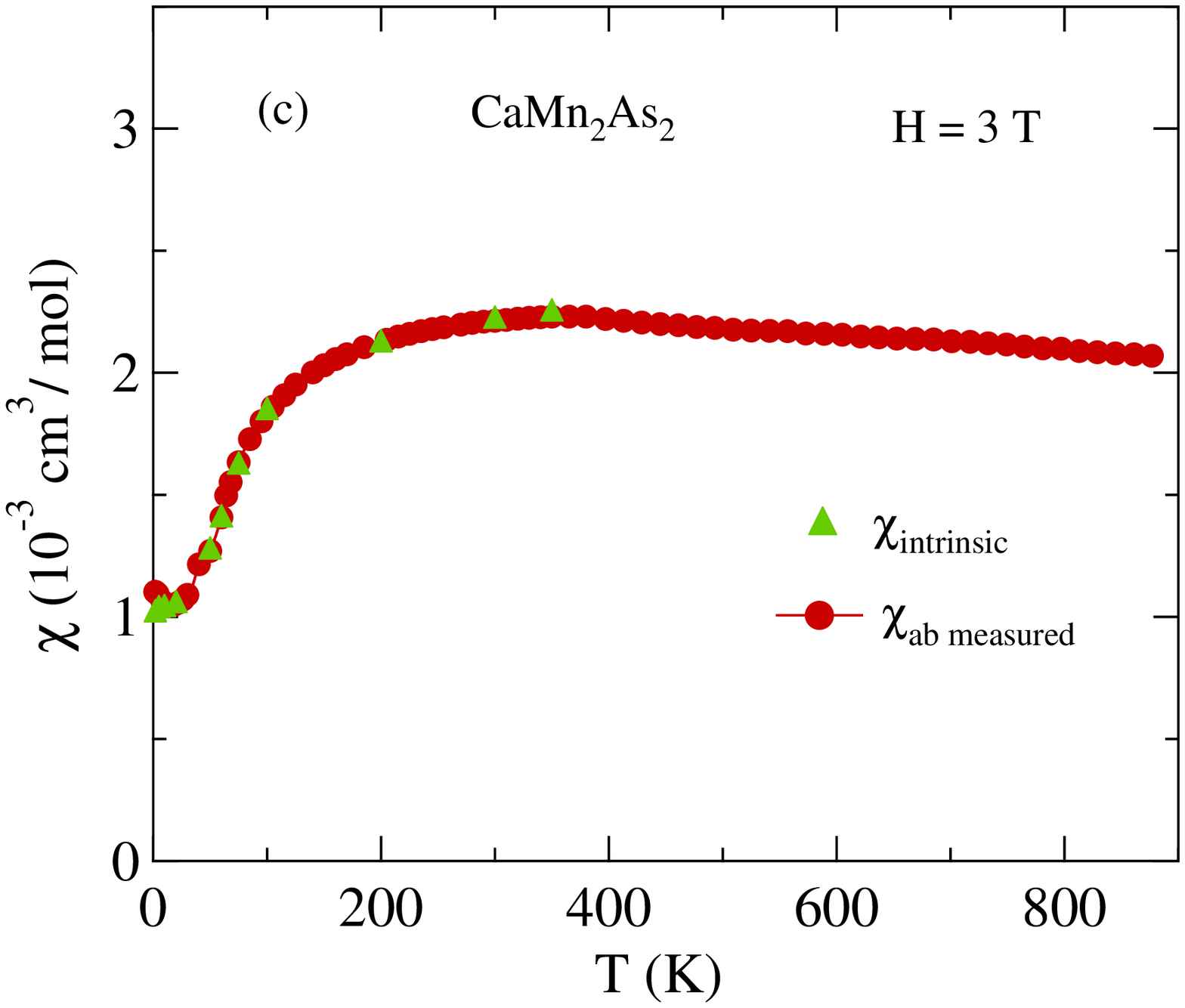}
\caption{(Color online) Magnetic susceptibility data for \cma{}.  The plot and symbol designations are the same as for \sma\ in Fig.~\ref{Fig:SrMn2As2_Chi}.}
\label{Fig:CaMn2As2_Chi}
\end{figure}

The zero-field-cooled (ZFC) magnetic susceptibility $\chi \equiv M/H$ versus~$T$ measured in $H = 0.1$~T and $H = 3$~T applied in the $ab$~plane ($H\parallel ab,\ \chi_{ab}$) and along the $c$~axis ($H\parallel c,\ \chi_c$) for single crystals of \sma{} and \cma{} are shown in Figs.~\ref{Fig:SrMn2As2_Chi} and~\ref{Fig:CaMn2As2_Chi}, respectively.  

Clear AFM transitions are observed in $\chi(T)$ at $T{\rm_N}\approx 120$~K for \sma{} and $T{\rm_N}\approx 65$~K for \cma{}, as indicated by  vertical arrows in Figs.~\ref{Fig:SrMn2As2_Chi}(a) and \ref{Fig:CaMn2As2_Chi}(a), respectively. We also performed FC (field-cooled) and ZFC $\chi(T)$ measurements at $H=0.1$~T and $H =3$~T (not shown here). No hysteresis was observed between the ZFC and FC data, which is consistent with long-range AFM ordering of \scma{} below their respective N\'eel temperatures.  The data in Figs.~\ref{Fig:SrMn2As2_Chi}(a) and \ref{Fig:CaMn2As2_Chi}(a) for $T>T_{\rm N}$ are nearly isotropic, as expected for Mn$^{+2}$ with spin $S=5/2$ and $g\approx2$.

From Figs.~\ref{Fig:SrMn2As2_Chi}(a) and \ref{Fig:CaMn2As2_Chi}(a), the anisotropy in $\chi$ at $T<T{\rm_N}$ indicates that the hard axis is the $c$~axis and the $ab$ plane is the easy plane for both compounds.  Furthermore, the nonzero limits of $\chi_{ab}(T\to0)$ suggest that the AFM structure could be either a collinear AFM with multiple domains aligned within the $ab$~plane or an intrinsic noncollinear structure with moments again aligned in the $ab$~plane.\cite{Johnston2012, Johnston2015, Anand2015, Ryan2015}  For collinear ordering, magnetic dipole interactions between the Mn moments favor $ab$-plane moment alignment over $c$-axis alignment.\cite{Johnston2016}  In \sma{} [see Fig.~\ref{Fig:SrMn2As2_Chi}(b)], the anisotropy in $\chi$ for $T<T{\rm_N}$ is eliminated by a field of 3~T, which indicates a relatively small magnetocrystalline anisotropy compared to that in \cma\ judging from Fig.~\ref{Fig:CaMn2As2_Chi}(b).  The small upturns in $\chi$ in Figs.~\ref{Fig:SrMn2As2_Chi} and~\ref{Fig:CaMn2As2_Chi} below $\sim 20$~K are believed due to trace amounts of paramagnetic local-moment impurities.

%___________________SrMn2As2_MH______________________

\begin{figure}
\includegraphics[width=3.5in]{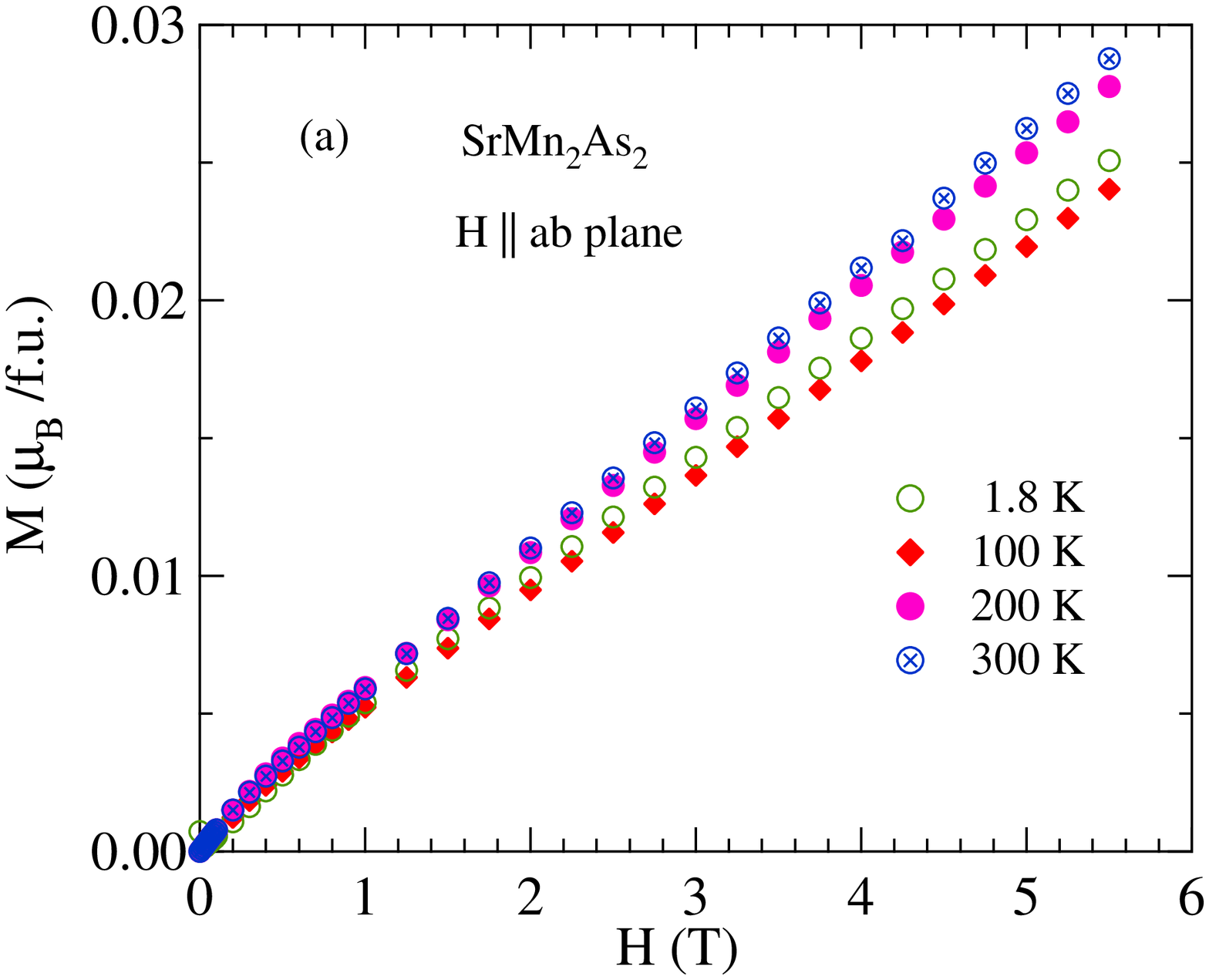}
\includegraphics[width=3.5in]{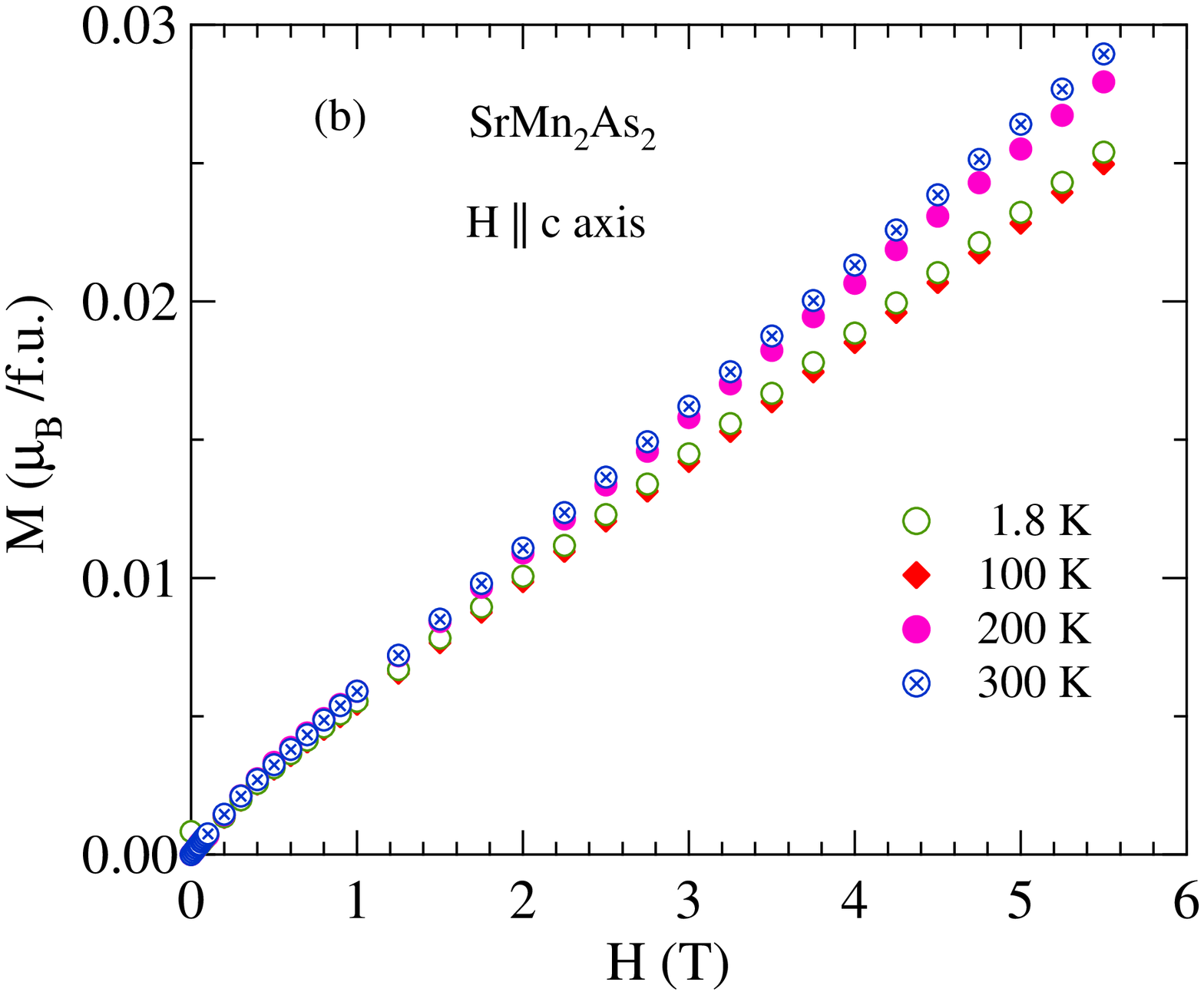}
 \protect\caption{(Color online) Magnetization $M$ versus magnetic field $H$ at various temperatures $T$ with (a) $H$ in the $ab$~plane ($H\parallel ab$) and with (b) $H$ along the $c$~axis ($H\parallel c$) for \sma{}.}
\label{Fig:SrMn2As2_MH}
\end{figure}
%-------------------------------------------------------------------------------------

%_________________________________________CaMn2As2_MH______________________
\begin{figure}
\includegraphics[width=3.5in]{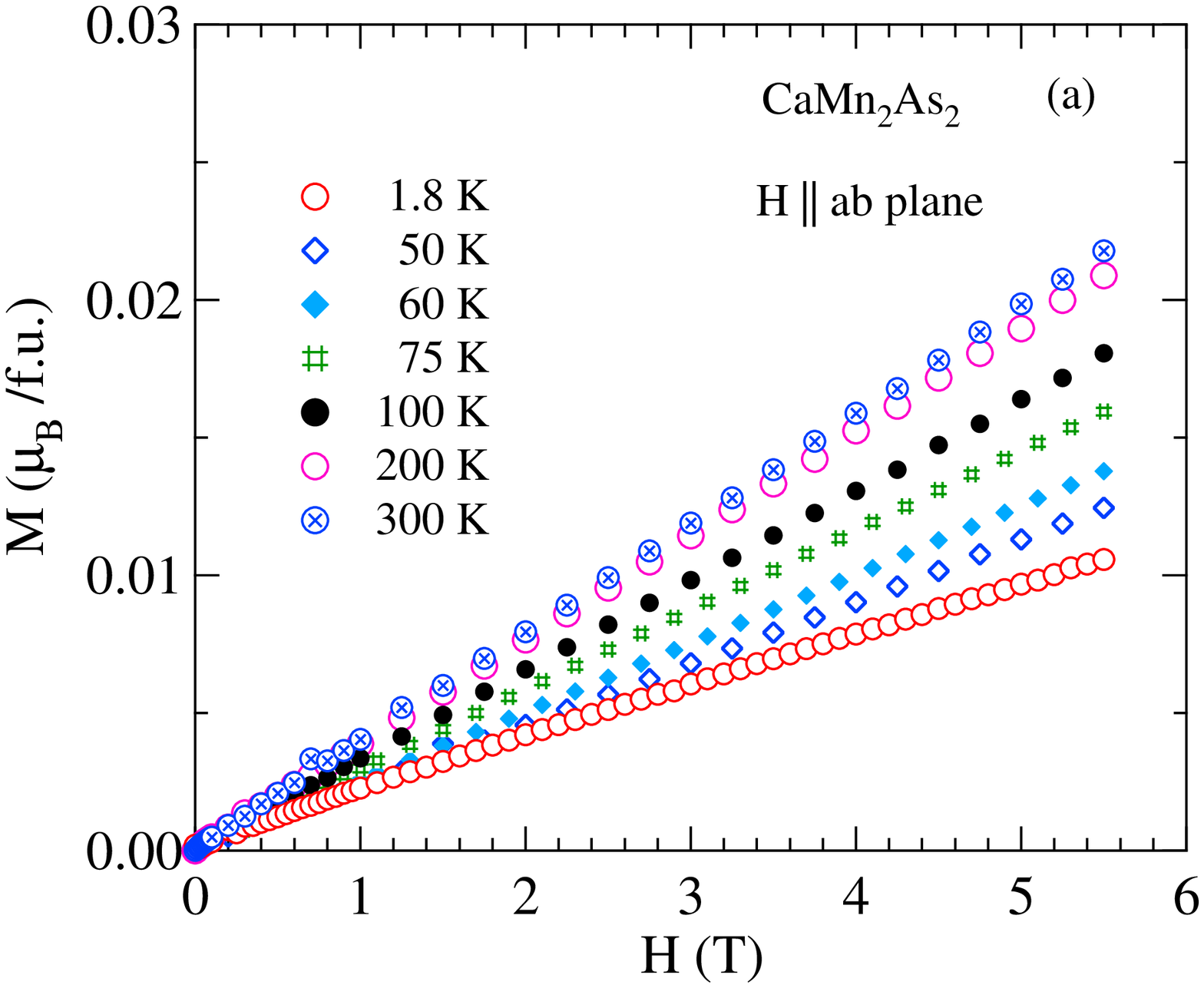}
\includegraphics[width=3.5in]{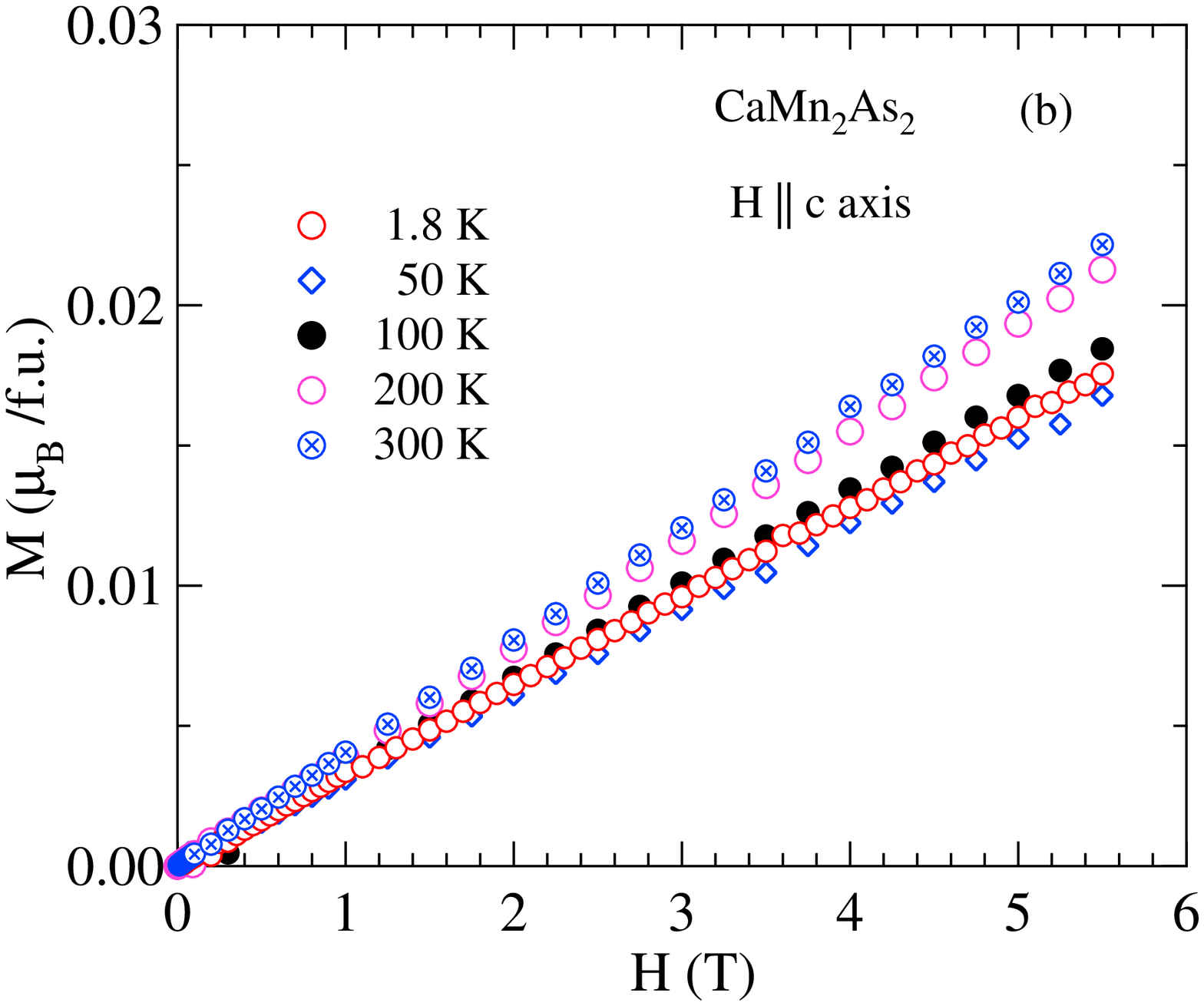}
 \protect\caption{(Color online) Magnetization $M$ versus magnetic field $H$ at various temperatures $T$ with (a) $H$ in the $ab$~plane ($H\parallel ab$) and with (b) $H$ along the $c$~axis ($H\parallel c$) for \cma{}.}
\label{Fig:CaMn2As2_MH}
\end{figure}
%----------------------------------------------------------------------------------------------------------

A small jump in $\chi(T)$ is observed for \sma{} in Fig.~\ref{Fig:SrMn2As2_Chi}(b) on cooling below about 320~K\@.  This is believed due to FM MnAs impurities with this Curie temperature that are present on the crystal surface and/or as an inclusion in the crystal, as previously observed for ${\rm BaMn_2As_2}$ crystals.\cite{Singh2009}  To extract the intrinsic magnetic behavior of \sma{}, we carried out $M(H)$ isotherm measurements at various temperatures. Figures~\ref{Fig:SrMn2As2_MH} and~\ref{Fig:CaMn2As2_MH} show isothermal $M(H)$ data for \sma{} and \cma{} crystals, respectively, at different $T$ for $H\parallel{ab}$ plane ($M_{ab}$) and $H\parallel{c}$ axis ($M_c$). The $M(H)$ curves are almost linear at high fields, but for \sma\ one sees nonlinearities at low fields ($H<1$--2 T) for $T<200$~K, confirming the presence of saturable FM impurities. 

To extract the intrinsic $\chi$ ($\chi_{\rm int}$) we fitted the $M(H)$ data in the high-field range $H=3.5$--5.5~T by the linear relation
\begin{equation}
M(H,T) = M_{\rm s}(T) + \chi_{\rm int}(T) H,
\label{eq:1}
\end{equation}
where $M_{\rm s}(T)$ is the saturation magnetization due to the FM impurities.  The $T\to0$ value of $M_{\rm s}$ for \sma\ is ${\rm 7~G~cm^3/mol} = 0.0013~\mu_{\rm B}$/f.u., which corresponds to 0.04~mol\% of MnAs impurities using the saturation moment of $\approx 3.5~\mu_{\rm B}$/f.u.\ (Refs.~\onlinecite{Haneda1977, Saparov2012}) for MnAs.  The $\chi(T)\equiv M(T)/H$ data in Figs.~\ref{Fig:SrMn2As2_Chi}(b) and~\ref{Fig:CaMn2As2_Chi}(b) were measured with $H=3$~T\@. Therefore,  we obtained the intrinsic $\chi$ from the isotherm data according to
\begin{equation}
\chi_{\rm int}(T)=\frac{M{\rm_{measured}}(T)-M{\rm_s}(T)}{3~\rm{T}}.
\label{Eq:Chicorrected}
\end{equation}
The $\chi_{\rm int}(T)$ data are shown by the filled green triangles in Figs.~\ref{Fig:SrMn2As2_Chi}(b) and~\ref{Fig:CaMn2As2_Chi}(b). It is seen that the $\chi_{\rm int}(T)$ data for \cma\ match very well for both field directions with the $\chi\equiv M/H$ data in Fig.~\ref{Fig:CaMn2As2_Chi}(b), indicating a clean crystal without any detectable FM impurities.

In order to further clarify the magnetism in these systems we measured $\chi(T)\equiv M(T)/H$ in the extended temperature range up to 900~K for \scma{} as shown in Figs.~\ref{Fig:SrMn2As2_Chi}(c) and \ref{Fig:CaMn2As2_Chi}(c), respectively. One sees that $\chi$ exhibits very broad maxima at $\sim 400$~K for both compounds. This feature is a signature of a low-dimensional local-moment AFM system.\cite{Johnston2011}  Thus \scma{} undergo a phase transition to a long-range ordered state below $T{\rm_N}$, preceeded by strong short-range AFM order at higher temperatures.  Indeed, the Curie-Weiss temperature region of $\chi$ is not reached even at 900~K, indicating that strong AFM correlations survive to significantly higher temperatures.

From the Mn--Mn interatomic distances discussed in Sec.~\ref{Sec:Struct}, we inferred that the Mn--Mn exchange coupling along the $c$~axis between the corrugated honeycomb Mn layers in the $ab$~plane is much smaller than within the layers.  We confirm this here and in addition infer that the Mn--Mn exchange coupling within the corrugated honeycomb layers is dominantly antiferromagnetic.

\subsection{Heat Capacity}

\begin{figure}[t]
 \includegraphics[width=3.5in]{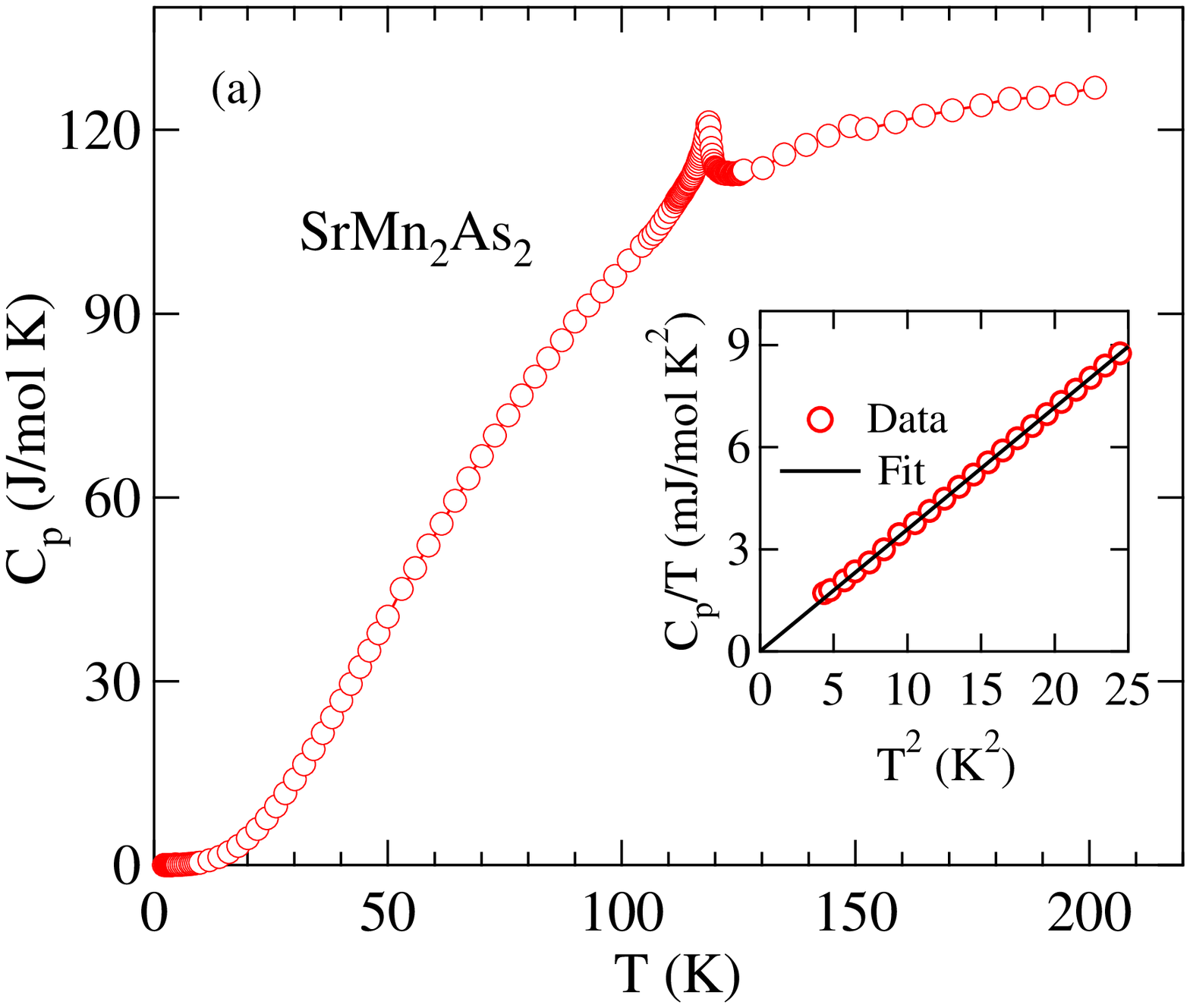}
 \includegraphics[width=3.5in]{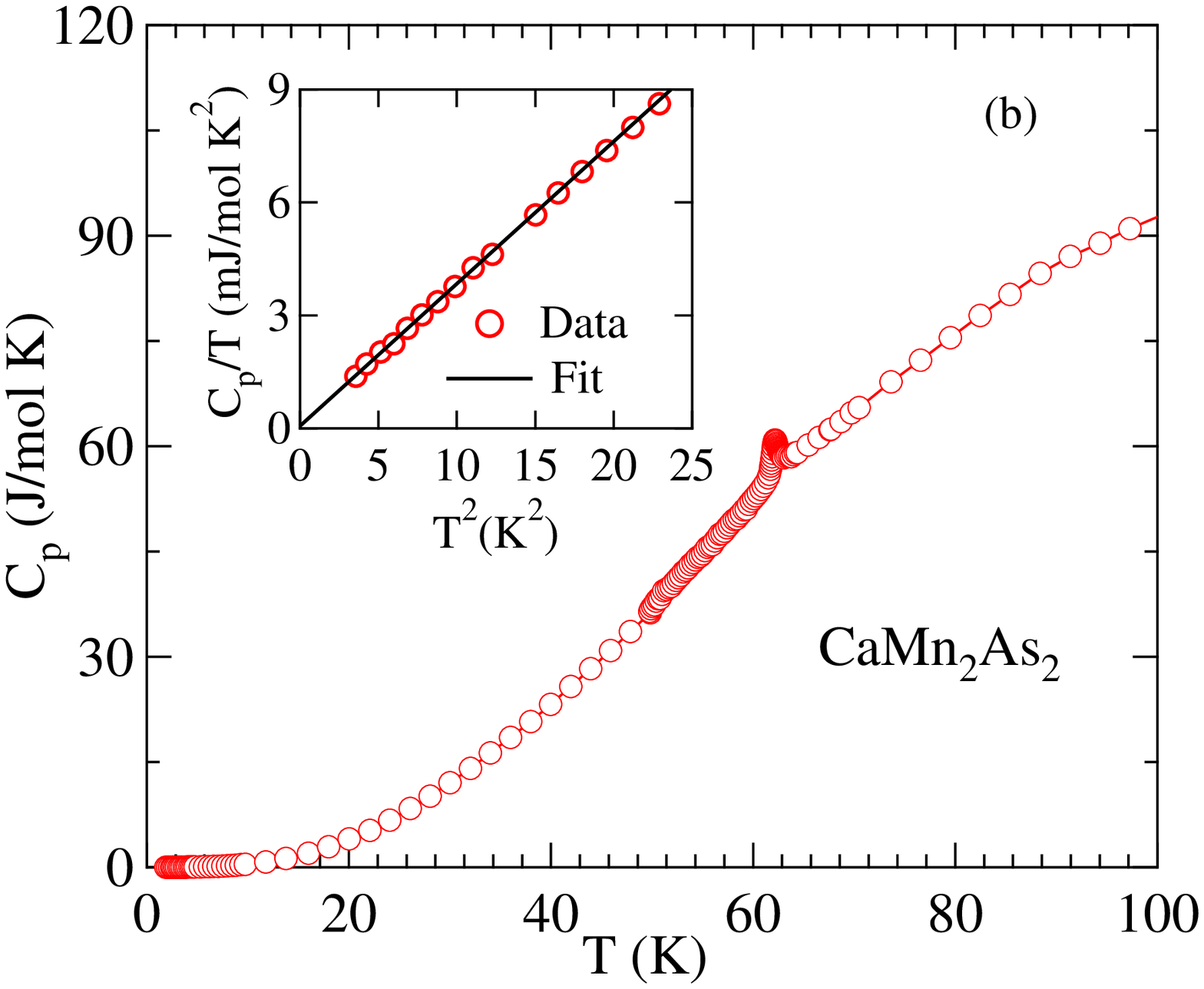}
\caption{(Color online) Heat capacity $C{\rm_p}$ versus temperature $T$ for (a) \sma{} and (b) \cma{}. The insets show $C{\rm_p}(T)/T$ versus $T^{2}$ for $T< 5$~K, where the straight lines though the respective data are fits by Eq.~(\ref{Eq:ConT}). }
\label{Fig:(Sr,Ca)Mn2As2_Cp}
\end{figure}

Figures~\ref{Fig:(Sr,Ca)Mn2As2_Cp}(a) and \ref{Fig:(Sr,Ca)Mn2As2_Cp}(b) show zero-field $C{\rm_p}(T)$ data for \sma{} and \cma{}, respectively. The sharp anomalies in $C_{\rm p}(T)$  at 120(2)~K in \sma{} and at 62(3)~K in \cma{} are the respective N\'eel temperatures of the two compounds, which are in good agreement with $T{\rm_N}$ values found above from the respective $\chi(T)$ data.

\subsubsection{Low-Temperature Behaviors}

The insets of Figs.~\ref{Fig:(Sr,Ca)Mn2As2_Cp}(a) and~\ref{Fig:(Sr,Ca)Mn2As2_Cp}(b) show $C{\rm_p}(T)/T$ versus $T^{2}$ between 1.8 and 5~K\@. At low temperatures we model the $C_{\rm p}(T)$ data by\cite{Kittel2005}
\be
C_{\rm p} = \gamma T + \beta T^3,
\ee
where the coefficient $\gamma$ is usually due to the electronic contribution (Sommerfeld coefficient) and $\beta$ is the coefficient of the Debye~$T^3$ lattice contribution in the absence of three-dimensional AFM spin-wave contributions. The data were therefore fitted by the expression
\be
\frac{C{\rm_p}}{T}=\gamma+\beta T^{2},
\label{Eq:ConT}
\ee
From the fits of Eq.~(\ref{Eq_thetaD}) to the data in the insets of Figs.~\ref{Fig:(Sr,Ca)Mn2As2_Cp}(a) and \ref{Fig:(Sr,Ca)Mn2As2_Cp}(b) we obtain $\gamma=0.0(1)$~mJ/(mol~K$^{2}$) for \sma{} and 0.05(7)~mJ/(mol K$^{2}$) for \cma{}. The null values of $\gamma$ are consistent with the insulating ground states found from the $\rho(T)$ measurements in Sec.~\ref{Sec:Rho}.

The fitted values for $\beta$ are
\be
\beta = 0.35(1)\,{\rm \frac{mJ}{mol\,K^4}}\ {\rm for\ SrMn_2As_2}
\label{betaForSMA}
\ee
and 0.37(1)~mJ/(mol~K$^{4})$ for \cma{}. 
We estimate the Debye temperatures $\Theta_{\rm D}$ for the two compounds from the Debye theory according to\cite{Kittel2005}
\be
\Theta{\rm_D}=\left(\frac{12\pi^{4}Rn}{5\beta}\right)^{1/3}
\label{Eq_thetaD}
\ee
where $R$ is the molar gas constant and $n$ is the number of atoms per formula unit [$n=5$ for \scma{}].  We obtain
\bse
\bea
\Theta{\rm_D} &=& 303(3)~{\rm K} \quad ({\rm SrMn_2As_2}),\label{Eq:QC_SMA}\\
 &=&297(3)~{\rm K} \quad ({\rm CaMn_2As_2}).
\eea
\ese
In the absence of anisotropy gaps in the AFM spin-wave spectrum, contributions to $\beta$ could arise from excitations of three-dimensional AFM spin waves at the low temperatures at which the $\beta$ values were extracted; hence the quoted values of $\Theta_{\rm D}$ are lower limits.

\subsubsection{Magnetic Contributions to the Heat Capacity and Entropy of \sma}

\begin{figure}[t]
\includegraphics[width=3in]{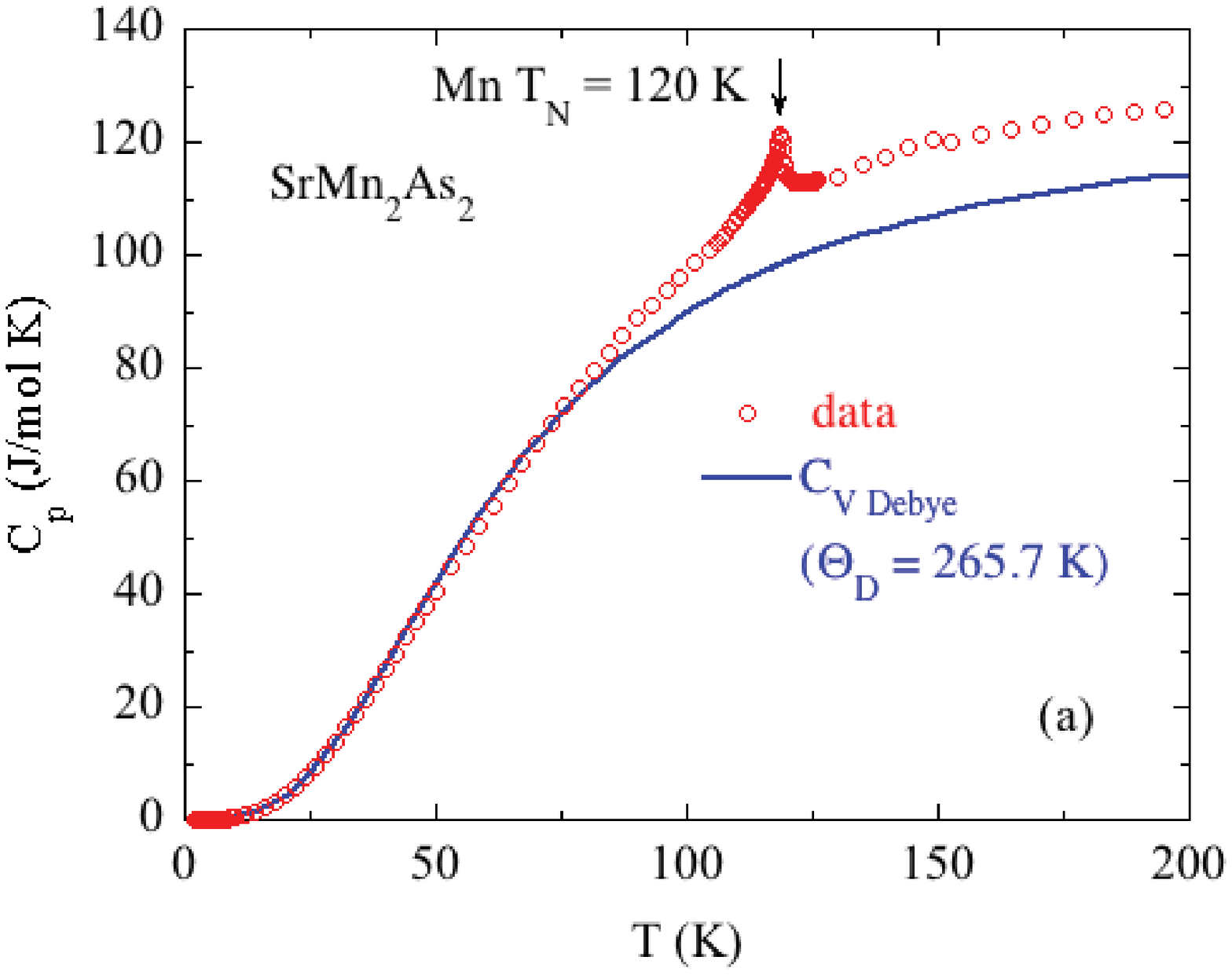}
\includegraphics[width=3in]{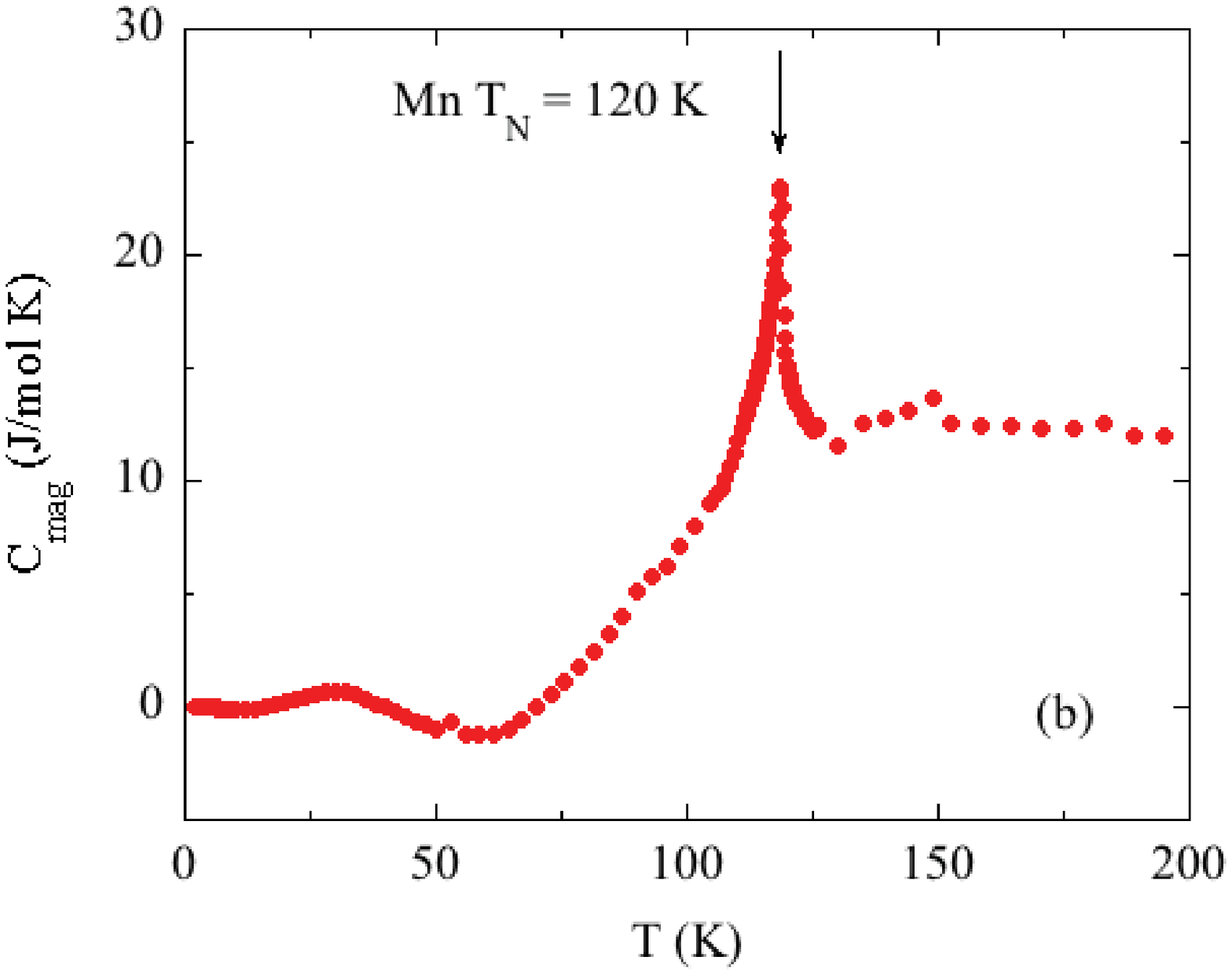}
\includegraphics[width=3in]{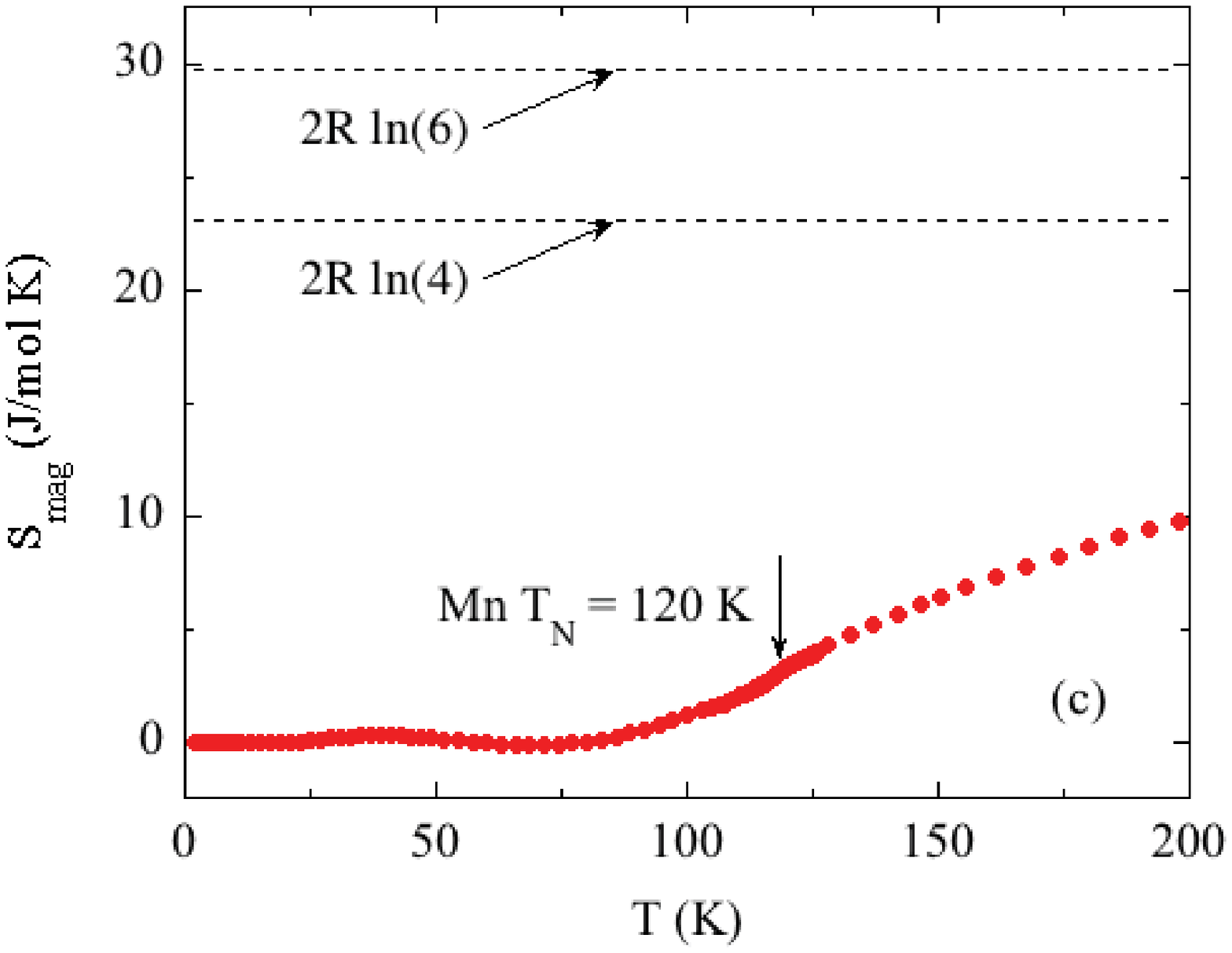}\vspace{-0.05in}
\caption{(Color online) (a)~Heat capacity $C{\rm_p}$ versus $T$ for \sma{} and a fit of $C_{\rm V\,Debye}(T)$ in Eqs.~(\ref{Eqs:HiTFit}) to the data for $T=1.8$~K to 60~K\@. (b)~Magnetic heat capacity $C_{\rm mag}(T)$. (c)~Magnetic entropy $S_{\rm mag}(T)$ obtained using Eq.~(\ref{Eq:Smag}).  The dashed lines are $S_{\rm mag}(T\to\infty)$ for $S=5/2$ and $S=3/2$.}
\label{Fig:SrMn2As2_Cp_CD}
\end{figure}

Here we extract estimates of the magnetic contribution $C_{\rm mag}(T)$ to the measured heat capacity of \sma\ and from that we obtain the magnetic contribution $S_{\rm mag}(T)$ to the entropy over the temperature range from 1.8~ to 200~K of the $C_{\rm p}(T)$ measurements. To accomplish this goal we first obtain an estimate of the lattice contribution $C_{\rm latt}(T)$ to the measured $C_{\rm p}(T)$.  We use the expression
\bse
\label{Eqs:HiTFit}
\be
C_{\rm latt}= n C_{\rm V\,Debye},
\ee
where $C_{\rm V\,Debye}$ is the Debye lattice heat capacity per mole of atoms given by\cite{Kittel2005}
\be
C_{\rm V\,Debye} = 9 R\left(\frac{T}{\Theta_{\rm D}}\right)^{3}\int_{0}^{\Theta_{\rm D}/T}\frac{x^{4}e^x}{\left(e^x-1\right)^2 }dx.
\label{Eq:Debye}
\ee
\ese
The representation of $C_{\rm V\,Debye}(T/\Theta_{\rm D})$ used here is an accurate analytic Pad\'e approximant function of $T/\Theta_{\rm D}$ obtained by fitting numerical solutions of Eq.~(\ref{Eq:Debye}) for a list of $T/\Theta_{\rm D}$ values.\cite{Goetsch2012}  The fit of $C_{\rm p}(T)$ for \sma\ in Fig.~\ref{Fig:SrMn2As2_Cp_CD}(a) by Eqs.~(\ref{Eqs:HiTFit}) over the temperature range from 1.8 to 60~K and its extrapolation is shown by the blue curve in Fig.~\ref{Fig:SrMn2As2_Cp_CD}(a) using the fitted Debye temperature $\Theta_{\rm D} = 265.7$~K\@.  This value of $\Theta_{\rm D}$ is comparable with the value of 303~K obtained from the fit to the $C_{\rm p}$ data for \sma\ at low~$T$ in Eq.~(\ref{Eq:QC_SMA}), especially considering that $\Theta_{\rm D}$ for a compound typically varies by $\pm20$\% on cooling from 300~K to 2~K.\cite{Smart1966}

The $C_{\rm mag}(T)$ is calculated as the difference between the measured $C_{\rm p}(T)$ and the fitted $C_{\rm latt}(T)$ in Fig.~\ref{Fig:SrMn2As2_Cp_CD}(a).  The result in shown in Fig.~\ref{Fig:SrMn2As2_Cp_CD}(b), where a sharp peak at $T_{\rm N}=120$~K is seen.  The $C_{\rm mag}$ at $T>T_{\rm N}$ shows that there is strong  dynamic short-range AFM order above $T_{\rm N}$\@.  The magnetic entropy $S_{\rm mag}(T)$ is calculated from $C_{\rm mag}(T)$ using
\be
S_{\rm mag}(T) = \int_0^T \frac{C_{\rm mag}(T)}{T}\,dT,
\label{Eq:Smag}
\ee
and the result is shown in Fig.~\ref{Fig:SrMn2As2_Cp_CD}(c).  The entropy of completely disordered spins~$S$ per mole of \scma\ is $S(T\to\infty) = 2R\ln(2S+1)$, which gives
\bse
\label{Eqs:SmagVals}
\bea
S_{\rm mag}(T\to\infty) &=& 23.1 \,{\rm \frac{J}{mol\,K}}\quad (S=3/2)\\*
&=& 29.8\,{\rm \frac{J}{mol\,K}}\quad (S=5/2),
\eea
\ese
as shown by the horizontal dashed black lines in Fig.~\ref{Fig:SrMn2As2_Cp_CD}(c).  This range of spin values encompasses the known variations in the ordered moments of Mn spins in various materials similar to ours which can arise from quantum fluctuation and/or hybridization effects (see, e.g., Ref.~\onlinecite{Johnston2011}).  We find that $S_{\rm mag}(200~{\rm K}) \approx 10~{\rm J/mol\,K}$ in Fig.~\ref{Fig:SrMn2As2_Cp_CD}(c) is only $\approx 33$\% of the value for $S=5/2$ in Eqs.~(\ref{Eqs:SmagVals}) and is still only $\approx 43$\% of the value for $S=3/2$. Thus the strong short-range AFM order revealed in the $C_{\rm mag}(T)$ and $S_{\rm mag}(T)$ data above $T_{\rm N}$ is consistent with the above conclusion from the $\chi(T)$ data that strong short-range AFM order survives from $T_{\rm N}$ up to at least 900~K\@. 

%\newpage

\section{Summary}

We have shown that \sma\ and \cma\ are AFM insulators with N\'eel temperatures $T_{\rm N} = 120$~K and 62~K, respectively. The microscopic origin of this large difference in magnitude of the N\'eel temperatures together with the reason why the Sr compound has a higher $T_{\rm N}$ than the Ca one, in spite of the smaller unit cell of the latter, remain to be explained.

The $\chi(T)$ data at $T\leq T_{\rm N}$ indicate that the hexagonal $c$~axis is a hard axis, with the ordered Mn spin-5/2 moments lying within the $ab$~plane. Since a collinear AFM structure within the $ab$~plane is inferred for \sma\ from neutron diffraction measurements that were carried out in a companion study,\cite{Das2016} the nonzero limits of $\chi_{ab}(T < T_{\rm N})$ observed for this compound must arise from the three collinear AFM domains with their axes at $60^\circ$ to each other within the $ab$~plane.  If the populations of the three domains are equal, within molecular field theory one obtains $\chi_{ab}(T\to0) = \chi(T_{\rm N})/2$, in approximate agreement with the $\chi_{ab}(T)$ data in Fig.~\ref{Fig:SrMn2As2_Chi}(a).  The prediction of the easy axis arising from the Mn--Mn magnetic dipole interactions in a collinear magnetic structure of \sma\ obtained using the formalism of Ref.~\onlinecite{Johnston2016} and the experimental crystal structure is that the ordered moments should lie in the $ab$~plane as inferred here from the $\chi(T)$ data and also observed\cite{Das2016} in the neutron diffraction experiments.

Thus the potential geometric frustration for AFM ordering within the triangular-lattice bilayers parallel to the $ab$~plane that originally motivated this work is apparently not important in \sma\ and \cma.  In particular, if AFM Mn--Mn exchange interactions within a triangular sublattice layer were dominant, a noncollinear AFM structure would have resulted instead of the observed\cite{Das2016} collinear AFM structure.

Strong dynamic AFM short-range correlations up to at least 900~K as observed in our $\chi(T)$ measurements, consistent with our $S_{\rm mag}(T)$ data up to 200~K, are likely due to quasi-two-dimensional connectivity of strong AFM Mn--Mn exchange interactions within the corrugated honeycomb Mn layers.  This in turn offers the possibility of novel electronic ground states arising upon doping these materials into the metallic state.

\acknowledgments

We thank P. Das, A. Kreyssig and A. I. Goldman for helpful discussions.  This research was supported by the U.S. Department of Energy, Office of Basic Energy Sciences, Division of Materials Sciences and Engineering.  Ames Laboratory is operated for the U.S. Department of Energy by Iowa State University under Contract No.~DE-AC02-07CH11358.

\end{document}